\documentclass[aps,prl,twocolumn,superscriptaddress,nofootinbib]{revtex4-1}
\usepackage[utf8]{inputenc}
\usepackage{amsmath}
\usepackage{amssymb}
\usepackage{epsfig}
\usepackage{graphicx,url}
\usepackage{bm}
\usepackage{mathrsfs}
\usepackage{color}
\usepackage[top=2cm, bottom=2.5cm, left=2cm, right=2cm]{geometry}
\usepackage{mathtools}
\usepackage{empheq}
\usepackage{physics}
\usepackage{xcolor}
\usepackage{ctable} 
\usepackage{array}
\usepackage{multirow}
\usepackage{subfig}
\usepackage{zref-base}
\usepackage{atveryend}
\makeatletter
\AfterLastShipout{%
  \if@filesw   
    \begingroup
      \zref@setcurrent{default}{\the\value{equation}}%
      \zref@wrapper@immediate{%
        \zref@labelbyprops{eq-last}{default}%
      }%
    \endgroup
  \fi
}
\makeatother

\newcolumntype{?}{!{\vrule width 1pt}}

\newcommand{\mb}{\mathbf}
\newcommand{\la}{\langle}
\newcommand{\ra}{\rangle}
\newcommand{\p}{\partial}
\newcommand{\f}{\frac}

\newcommand{\mr}{\mathrm}

\newcommand{\nn}{\nonumber}

\begin{document}

\title{Reconsidering the structure of nucleation theories}

\author{Anja Kuhnhold}
\affiliation{\it Physikalisches Institut, Albert-Ludwigs-Universit\"{a}t,  79104 Freiburg, Germany}
\author{Hugues Meyer}
\affiliation{\it Physikalisches Institut, Albert-Ludwigs-Universit\"{a}t,  79104 Freiburg, Germany}
\affiliation{Research Unit in Engineering Science, Universit\'{e} du Luxembourg,\\  L-4364 Esch-sur-Alzette, Luxembourg}
\author{Graziano Amati}
\affiliation{\it Physikalisches Institut, Albert-Ludwigs-Universit\"{a}t,  79104 Freiburg, Germany}
\author{Philipp Pelagejcev}
	\affiliation{\it Physikalisches Institut, Albert-Ludwigs-Universit\"{a}t,  79104 Freiburg, Germany}
\author{Tanja Schilling}
\affiliation{\it Physikalisches Institut, Albert-Ludwigs-Universit\"{a}t,  79104 Freiburg, Germany}

\date{\today}

\begin{abstract}
We discuss the structure of the equation of motion that governs nucleation processes at first order phase transitions. From the underlying microscopic dynamics of a nucleating system, we derive by means of a non-equilibrium projection operator formalism the equation of motion for the size distribution of the nuclei. The equation is exact, i.e.~the derivation does not contain approximations. To assess the impact of memory, we express the equation of motion in a form that allows for direct comparison to the Markovian limit. As a numerical test, we have simulated crystal nucleation from a supersaturated melt of particles interacting via a Lennard-Jones potential. The simulation data show effects of non-Markovian dynamics.
\end{abstract}

\maketitle

\section*{Introduction}

Nucleation is part of a broad class of physical processes which are described in terms of ``reaction coordinates'', i.e.~processes for which it is useful to reduce the description of the complex microscopic dynamics to a small set of observables that capture the essential features.  Nucleation phenomena have impact in diverse scientific fields \cite{oxtoby:1992, kelton:2010}. If, for instance, a metal melt is cooled to solidify, the mechanical properties of the product will depend on details of the cooling process and, in particular, on the rate at which crystallites nucleate and grow \cite{finney:2008, kashchiev:2000}. Similarly, in the atmosphere liquid droplets or crystallites nucleate from supercooled water vapour \cite{cantrell:2005, viisanen:1993, zhang:2011}. The details of the size distribution and morphology of these aggregates have an impact on the weather.

The common feature of all nucleation processes is that a system is initialized in a metastable state and is expected to reach a qualitatively different, stable state in the long-time limit after crossing a first order phase transition. Although the process involves a very large number of microscopic degrees of freedom, the standard way of describing it focusses on the dynamics of a simple reaction coordinate, in most cases the size of a droplet\footnote{We will use the term ``droplet'' throughout this article, but our arguments apply equally to aggregates that precipitate from solution and crystallites that form in a supercooled melt.} (aggregate, cluster or crystallite, resp.).  ``Classical Nucleation Theory'' (CNT) is the prevalent theoretical approach used to analyze the dynamics of this reaction coordinate \cite{volmer:1926, becker:1935, kalikmanov:2013}.
The main idea underlying CNT is to assume that the probability of forming a droplet of a certain size is governed by the interplay between a favourable volume term, driven by the chemical potential difference between the metastable phase and the stable phase, and an unfavourable interfacial term controlled by the interfacial tension. The competition between these opposite contributions produces a free energy barrier that can be overcome due to thermal fluctuation. These concepts are accompanied by an additional assumption:
the evolution is expected to be Markovian, which allows to model the process by a memory-less Fokker-Planck equation of the form of eqn.~(\ref{standardFP}) where the drift term $a_{1}$ includes the free energy competition between volume and surface contributions.

Although this picture yields good qualitative results, it fails to reproduce experimental and numerical data quantitatively, often even by many orders of magnitude \cite{gebauer:2011,hegg:2009,sear:2012,filion:2010,horsch:2008,tanaka:2005,russo:2012}. Explanations for these discrepancies have been offered on different levels: by considering inconsistencies in the functional form of the free energy (see e.g.~the review by Laaksonen and Oxtoby \cite{laaksonen:1995} or the one by Ford \cite{ford:2004}), by addressing the choice of reaction coordinate \cite{moroni:2005,peters:2006,barnes:2014,leines:2017,lechner:2011}, the infinite size of the system \cite{schweitzer:1988}, the fixed position of the droplet in space \cite{ford:1997}, the simple form of the free energy profile which does not account for the structure of the droplet \cite{omalley:2003, sanz:2007, binder:2016}, by including nonclassical effects in a density-functional approach \cite{oxtoby:1988,oxtoby:1998, prestipino:2012}, or by using dynamical density-functional-theory instead of the over-simplified free energy picture \cite{lutsko:2012,lutsko:2013}, by testing the capillarity approximation \cite{schrader:2009}, by adapting the value of the interfacial tension\footnote{To ``correct'' the value of the interfacial tension in retrospect in order to make CNT predictions fit the experimental data is such a common strategy, that we would need to list hundreds of references here.} and by challenging the basic assumptions of transition state theory, i.e.~the accuracy of the Markovian approximation \cite{jungblut:2015} and the validity of a Fokker-Planck description \cite{shizgal:1989, sorokin:2017,kuipers:2010}. We will discuss in this article in particular the latter point, non-Markovian effects.

A droplet of a certain size can be realized by a large number of different microscopic configurations. When modeling nucleation we do thus inevitably deal with a coarse-graining problem, i.e.~we reduce the description of the full microscopic problem to that of one quantity averaged over a non-equilibrium ensemble of microscopic trajectories. Often it is useful to model coarse-grained variables in a probabilistic way (although, in principle, one could derive a deterministic equation of motion for a coarse-grained variable from a bundle of underlying deterministic microscopic trajectories). A common strategy is to work on the level of the probability distribution $p(\alpha,t)$ of the observable $A$, that is the probability that the observable $A$ has the value $\alpha$ at time $t$. In cases of ergodic dynamics without external driving, $p(\alpha,t)$ is expected to reach an equilibrium distribution $p_{\beta}(\alpha)$ in the long-time limit. At all times, one can relate $p(\alpha,t)$ to the time-dependent phase-space probability density, $\rho(\mathbf{\Gamma},t)$, that corresponds to the ensemble of trajectories, via
\begin{equation}
p(\alpha,t) = \int \mr d\mathbf{\Gamma} \rho(\mathbf{\Gamma},t) \delta(\alpha - A(\mathbf{\Gamma}))
\end{equation}
If the dynamics of the coarse-grained variable is Markovian, the Fokker-Planck equation is sufficient to describe the dynamics of $p(\alpha,t)$, i.e. 
\begin{equation}
\label{standardFP}
\frac{\partial p(\alpha,t)}{\partial t} = \frac{\partial}{\partial \alpha} \left[ a_{1}(\alpha)p(\alpha,t)  \right] + \frac{\partial^{2}}{\partial \alpha^{2}} \left[ a_{2}(\alpha) p(\alpha,t)  \right]
\end{equation}
where $a_{1}$ and $a_{2}$ are called drift and diffusion coefficients, respectively. Although it is difficult to assess a priori whether a coarse-grained variable has Markovian dynamics, the Fokker-Planck equation is often used to analyse epxerimental or numerical results.

Here, we derive the structure of the full, non-Markovian, equation of motion of $p(\alpha,t)$. By applying a suitable projection operator to the underlying microscopic dynamics, we obtain the equation of motion that contains memory, takes the form of a non-local Kramers-Moyal expansion and allows us to draw a direct comparison to the Fokker-Planck equation. To illustrate the difference between the exact theory and the approximative Fokker-Planck description, we analyze crystal nucleation trajectories from molecular dynamics simulation and show that the evolution of the crystallite size distribution is non-Markovian.

\section*{Derivation of the Equation of Motion}

\subsection*{Reminder of Grabert's Approach}

Projection operator techniques are often used to derive Generalized Langevin Equations for a set of dynamical variables such as e.g.~the reaction coordinates of a complex process. These techniques are based on the definition of a projection operator that distinguishes a main contribution to the dynamics, the so-called drift term, from a marginal one. The choice of the projection operator can be adapted in order for the drift term to be tuned to the problem under study. Grabert has suggested how to use these techniques instead in order to derive an equation of motion for the probability density $p(\alpha,t)$ of a dynamical variable $A$, i.e.~the probability for the variable $A$ to be equal to $\alpha$ at time $t$ \cite{grabert:1982} (which is a description of the generalized Fokker-Planck-form rather than the Langevin-form).

Based on an arbitrary phase-space observable $A$, i.e.~a variable that is fully determined by the position $\mathbf{\Gamma}$ in phase-space, we define distributions $\psi_\alpha$ that act on states $\mb \Gamma$ as
\begin{equation}
\psi_\alpha (\mb \Gamma) = \delta(A(\mb \Gamma)-\alpha)
\end{equation}
These distributions are themselves completely determined by the position $\mathbf{\Gamma}$ in phase-space and can thus be treated as dynamical variables for which we can apply projection operator techniques. The following projection operator is then defined:
\begin{equation} 
\label{proj}
P X(\mb \Gamma) = \int \mr d \alpha \f{\int \mr d \mb \Gamma' \rho_{\beta}(\mb \Gamma') \psi_\alpha(\mb \Gamma') X(\mb \Gamma')}{p_{\beta}(\alpha)} \psi_\alpha(\mb \Gamma) 
\end{equation}
where $X$ is an arbitrary dynamical variable and
\begin{equation} \label{p_b_a}
p_{\beta}(\alpha) = \int \mr d \mb \Gamma \rho_{\beta}(\mb \Gamma) \psi_\alpha(\mb \Gamma) 
\end{equation}
is the equilibrium probability density corresponding to the dynamical variable $A$ and $\beta = 1/k_BT$. In words, $PX(\mathbf{\Gamma})$ is the sum of the equilibrium averages of the observable $X$ in all the subspaces $A(\mathbf{\Gamma})=\alpha$ weighted each with their equilibrium probability. It is easily verified that $P^{2} =P$, i.e.~$P$ is a projection operator. In particular, we have
\begin{equation}
P\left[f(\mathbf{\Gamma})\delta(\alpha - A(\mathbf{\Gamma}) \right] \propto \delta(\alpha - A(\mathbf{\Gamma}))
\end{equation}
for any phase space function $f(\mathbf{\Gamma})$.
Now we would like to obtain an equation of motion for $\psi_\alpha(t)$, the average of which is the out-of-equilibrium time-dependent probability distribution $p(\alpha,t)$ of $A$, namely
\begin{equation}
p(\alpha,t) \equiv \int \mr d \mb \Gamma \rho(\mb \Gamma, t)  \psi_\alpha(\mb \Gamma)
\end{equation}
where $\rho(\mb \Gamma, t)$ is the out-of-equilibrium phase-space density. As in any projection operator formalism, the main idea of the derivation is to split the propagator $e^{i\mathcal{L}t}$, where $i\mathcal{L}$ is the Liouville operator of the underlying microscopic model, into a parallel and an orthogonal contribution. The standard Dyson decomposition yields \cite{hansen:1990}
\begin{equation}
\label{Dyson}
e^{i\mathcal{L} t} =  e^{i\mathcal{L}t}P + \int_{0}^{t}{\mr d\tau e^{i\mathcal{L}\tau}P i\mathcal{L}  Q e^{i\mathcal{L}(t-\tau)Q} }  + Qe^{i\mathcal{L}tQ}
\end{equation}
where $Q=1-P$. Most of the following steps consist in mathematical transformations relying on the identity $i\mathcal{L}\psi_{\alpha}(\mb \Gamma) = -\psi_{\alpha}i\mathcal{L}A(\mb \Gamma)$ and on the fact that $\rho_{\beta}$ is the equilibrium phase-space density, which implies $i\mathcal{L}\rho_{\beta} = 0$ (for details see supplemental material, as well as ref.~\cite{grabert:1982}). The resulting equation of motion is 
\begin{align} 
\label{exact_fp}
\frac{\partial p(\alpha,t)}{\partial t} &= - \f{\p }{\p \alpha} \left[ w_{\beta}^{(1)}(\alpha)p(\alpha,t) \right] \nonumber \\
+ \int_{0}^{t} \mr d\tau &\frac{\partial}{\partial \alpha} \int \mr d\alpha' D(\alpha,\alpha',t-\tau)p_{\beta}(\alpha') \frac{\partial}{\partial \alpha'}\left( \frac{p(\alpha',\tau)}{p_{\beta}(\alpha')} \right)  
\end{align}
where 
\begin{align}
\label{def_w}
w_{\beta}^{(i_{1},\cdots,i_{p})}&(\alpha) =\frac{1}{p_{\beta}(\alpha)}\times\nonumber\\ \int \mr d\mb\Gamma& \rho_{\beta}(\mb\Gamma)\psi_{\alpha}(\mb\Gamma) \left[\left(i\mathcal{L}\right)^{i_{1}} A(\mb\Gamma)\right] \cdots \left[\left(i\mathcal{L}\right)^{i_{p}} A(\mb\Gamma)\right]    \\
\label{defD}
D(\alpha, \alpha',t) &= \frac 1 {p_\beta(\alpha')} \int  \mr d \mb \Gamma' \rho_{\beta}( \mb \Gamma')  R_{\alpha}(t, \mb \Gamma') R_{\alpha'}(0, \mb \Gamma') \\
\label{defR}
R_{\alpha}(t,\mb\Gamma) &= Qe^{i\mathcal{L}Qt}\psi_{\alpha}(\mb\Gamma)i\mathcal{L}A(\mb\Gamma)
\end{align}
Note that this equation is valid only if the last term of the Dyson decomposition eqn.~(\ref{Dyson}) vanishes. This holds if the initial phase-space density $\rho(\mb\Gamma,0)$ as well as the equilibrium density are so-called ``relevant densities'', i.e.~they are fully determined by the probability distributions $p(\alpha, 0)$ and $p_{\beta}(\alpha)$. Formally, this condition is written as:
\begin{equation}
\frac{\rho(\mb\Gamma,0)}{\rho_{\beta}(\mb\Gamma)} = \frac{p(A(\mb\Gamma), 0)}{p_{\beta}(A(\mb \Gamma))} 
\end{equation}
This condition implies that the observable $A$ must be chosen carefully: in the initial non-equilibrium state as well as in the final equilibrated one, all the microstates $\mathbf{\Gamma}$ such that $A(\mathbf{\Gamma}) = \alpha$ must be equivalent. 
 
\subsection*{Kramers-Moyal Expansion}
Now we consider the formation and growth of a droplet of the stable phase that emerges from a metastable bulk phase after a quench. A variable that measures the size of the droplet is a natural reaction coordinate for this process. However, we need to keep in mind that, in order for the formalism derived in the previous paragraph to apply, the variable $A$ must be fully determined by the position of the system in phase-space. There could be several droplets in one single system at the same time. Their size distribution would not be a variable of the type defined above, while e.g.~the size of the largest droplet in the system or the average size of all droplets present simultaneously would be suitable variables. The specific choice of the reaction coordinate will have an impact on the quantitative application of the theory, but the general structure of the resulting equations will not be affected. We will therefore develop our arguments under the assumption that the reaction coordinate is a variable $N$ that counts the number of particles in the largest droplet in the system. Note that we will change the notation $A$ and $\alpha$ to $N$ and $n$, respectively. 
 
Let us simplify eqn.~(\ref{exact_fp}), or at least cast it in a more intuitive form. Since our observable depends only on the positions of the particles (and not on their momenta), we can easily show that all functions $w^{(i_{1},\cdots,i_{p})}_{\beta}(n)$ vanish as long as $\sum_{k=1}^{p}i_{k}$ is an odd number. This result is a direct consequence of the invariance of the equilibrium phase-space density $\rho_{\beta}(\mathbf{\Gamma})$ under the transformation $\textbf{p}_{i} \rightarrow - \textbf{p}_{i}$, where $\textbf{p}_{i}$ is the momentum of the particle $i$. Thus, $w_{\beta}^{(1)}(n) = 0$, and the first term of eqn.~(\ref{exact_fp}) vanishes.

The second step is to recast $p_{\beta}(n)$ in terms that allow for a direct comparison between the theory we develop here and free-energy based theories such as CNT. In equilibrium, the probability of finding a certain macrostate can be related to an effective free energy. In particular, given the observable $N$ we can define a ``free energy profile'' $\Delta G(n)$ that is related to the probability $p_{\beta}(n)$ via
\begin{align}
\label{def_DG}
\Delta G(n) &:= -\frac{1}{\beta} \ln(p_{\beta}(n)) \nonumber \\
&= -\frac{1}{\beta} \ln \left[ \int \mr d\mathbf{\Gamma} \rho_{\beta}(\mathbf{\Gamma}) \delta(n - N(\mathbf{\Gamma})) \right] 
\end{align}
This definition is consistent with the notion of the free energy of a bulk equilibrium system, and it allows us to write $p_{\beta}(n) = e^{-\beta \Delta G(n)}$. Note, however, that we have not used any additional bulk, equilibrium observables as input such as e.g.~an interfacial tension or a supersaturation to define $\Delta G(n)$. In particular, we have not invoked the capillarity approximation.

 We can thus transform eqn.~(\ref{exact_fp}) noting that
\begin{equation}
p_{\beta}(n) \frac{\partial}{\partial n}\left( \frac{p(n,\tau)}{p_{\beta}(n)} \right)  = \left(\frac{\partial}{\partial n} + \beta\frac{\partial \Delta G (n)}{\partial n} \right)p(n,\tau)
\end{equation}

At this stage, eqn.~(\ref{exact_fp}) is still non-local in $n$, and our final goal is to obtain an equation that can be easily compared to the Fokker-Planck equation. We will therefore decompose the non-locality into a Kramers-Moyal expansion with memory. To do this, we first Taylor-expand the phase-space function $R_{n}(t,\mathbf{\Gamma})$ defined in eqn.~(\ref{defR}), i.e.
\begin{equation}
R_{n}(t,\mathbf{\Gamma}) = \sum_{p=0}^{\infty}\frac{t^{p}}{p!}Q\left[i\mathcal{L}Q\right]^{p}\psi_{n}(\mb\Gamma)i\mathcal{L}N(\mb\Gamma)
\end{equation}
Given the relation $i\mathcal{L}\psi_{n} = -\frac{\partial \psi_{n}}{\partial n} i\mathcal{L}N$ and that for any variable $X(\mathbf{\Gamma})$ we have $P\left[X\psi_{n}\right] \propto \psi_{n}$, we obtain the following structure
\begin{equation}
\label{struct_R}
\left.\frac{\partial ^{l} R_n}{\partial t^{l}}\right|_{n,t=0} = \sum_{k=0}^{l} r_{l,k}(n,\mathbf{\Gamma})\frac{\partial^{k}\psi_{n}}{\partial n^{k}}
\end{equation}
This identity is proven by induction in the supplemental material, and an expression for $r_{l+1,k}(n,\mathbf{\Gamma})$ is given in terms of all the preceding terms $r_{l'\leq l,k}(n,\mathbf{\Gamma})$.  

Inserting eqn.~(\ref{struct_R}) into eqn.~(\ref{defD}), we obtain after some algebra
\begin{equation}
\label{exp_D}
D(n,n',t) = \sum_{k=0}^{\infty}\frac{\partial^{k} }{\partial n^{k}} \left[  d_{k}(n,n',t) \delta(n-n')  \right]
\end{equation}
where the functions $d_{k}(n, n',t)$ are defined by
\begin{widetext}
\begin{equation}
\label{ea:defd}
d_{k}(n,n',t) =  \frac{1}{p_{\beta}(n')} \int \mr d\mathbf{\Gamma} \rho_{\beta}(\mathbf{\Gamma}) \left[ \sum_{p=k}^{\infty} \zeta_{p,k}(n,\mathbf{\Gamma}) \frac{t^{p}}{p!} \right] i\mathcal{L}N(\mathbf{\Gamma}) \delta(n-N(\mathbf{\Gamma})) 
\end{equation}
\end{widetext}
and 
\begin{equation}
\zeta_{p,k}(n) = \sum_{k'=k}^{p} (-1)^{k'-k}{{k'}\choose{k}} \frac{\partial^{k'-k}r_{p,k'}(n,\mathbf{\Gamma})} {\partial n^{k'-k}}
\end{equation}
We will then set $\tilde{d}_{k}(n,t) := d_{k}(n, n,t)$, the Taylor expansion of which can be expressed in terms of the functions $w_{\beta}^{(i_{1},\cdots,i_{p})}(n)$. 
The expansion eqn.~(\ref{exp_D}) serves to transform the non-locality in $n$ into a sum of contributions of all the derivatives of $p(n,t)$ with respect to $n$. The equation of motion of the time-dependent probability distribution of the droplet size then becomes
\begin{widetext}
\begin{equation}
\label{FP_dk}
\frac{\partial p(n, t)}{\partial t} =  \int_{0}^{t} \mr d\tau  \sum_{k=0}^{\infty}  \frac{\partial^{k+1}}{\partial n^{k+1}}  \left[\tilde{d}_{k}(n,t-\tau)  \left(  \frac{\partial }{\partial n}   + \beta \frac{\partial \Delta G(n)}{\partial n} \right) p(n,\tau)  \right]
\end{equation}
\end{widetext}
which is the central result of our work. Note that this expression is exact, i.e.~up to here the derivation did not contain any approximation.

The structure of eqn.~(\ref{FP_dk}) is similar to a Fokker-Planck equation but it differs from eqn.~(\ref{standardFP}) in two major aspects: the non-locality in time  and the sum involving an infinite number of effective diffusion constants $\tilde{d}_{k}$. The first aspect implies that nucleation dynamics will, in general, not be Markovian. The second aspect implies that a function $\tilde{d}_{l}$ modifies the evolution of the moments of $p(n,t)$ at orders larger than $l$. Thus, if the series cannot be truncated at order $k=1$, the evolution of $p(n,t)$  is not simply diffusive. One consequence of these two effects is that the definition of the term ``nucleation rate'' is not entirely obvious anymore. This might be one of the sources of the discrepancy between experimentally observed  and theoretically predicted nucleation rates.

Given the complexity of the terms $\tilde{d}_{k}$, we did not find simple estimates which would hold in general for all nucleation processes independently from the details of the microscopic dynamics and the preparation of the initial state.\footnote{This finding might be disappointing, but it agrees with the experimental observation, that CNT can be wrong by orders of magnitude in both directions \cite{gebauer:2011,hegg:2009,sear:2012,filion:2010,horsch:2008,tanaka:2005,russo:2012}.} However, we will lay out in the following section how eqn.~(\ref{FP_dk}) compares to existing nucleation theories, and we will illustrate the differences by means of computer simulation.

Before comparing eqn.~(\ref{FP_dk}) to CNT, we rewrite it as a non-Markovian Kramers-Moyal expansion \cite{risken:1996}
\begin{equation}
\label{KM}
\frac{\partial p(n, t)}{\partial t} =  \sum_{k=1}^{\infty}   \frac{\partial^{k}}{\partial n^{k}}  \int_{0}^{t} \mr d\tau   \left[\mathcal{D}^{(k)}(n,t-\tau)   p(n,\tau)  \right]
\end{equation}
where the coefficients $\mathcal{D}^{(k)}$ are identified as
\begin{equation}\label{D1}
\mathcal{D}^{(1)}(n,t) = \tilde{d}_{0}(n,t)\beta \frac{\partial \Delta \mathcal{G}_{\beta}^{(0)}(n, t)}{\partial n}
\end{equation}
and 
\begin{align}
\mathcal{D}^{(k)}(n,t) = & \tilde{d}_{k-1}(n,t)\beta \frac{\partial \Delta \mathcal{G}_{\beta}^{(k-1)}(n, t)}{\partial n} +  \tilde{d}_{k-2}(n,t)
\end{align}
for $k\geq 2$, where we have defined $\Delta \mathcal{G}_{\beta}^{(k)}(n, t) = \Delta G(n) - \ln\left(\tilde{d}_{k}(n,t) \right)/\beta$. This final recasting of the equation can be useful in order to evaluate the time-evolution of the moments of the distribution. 

\section*{Derivation of CNT}

In CNT (and most other approaches to nucleation that are based on a free energy landscape) the nucleation process is described by a standard Fokker-Planck equation, i.e
\begin{equation}
\label{FP_CNT}
\frac{\partial p(n, t)}{\partial t} =  \frac{\partial}{\partial n}  \left[D_{0}(n)  \left(  \frac{\partial }{\partial n}   + \beta  \Delta G'(n) \right) p(n,t)  \right]
\end{equation}
where $ \Delta G'(n) = \partial \Delta G(n) / \partial n $.
The functional form of the free energy profile has been, and still is, a subject of debate. There is consensus in the literature about the fact that $\Delta G(n)$ is determined by an interplay between a favourable drift term, which increases with the volume of the droplet and the thermodynamic driving force of the phase transition, and an unfavourable surface term controlled by the interfacial tension, and also about the fact that the competition between the terms creates a barrier that needs to be overcome in order for the stable phase to grow. However, details of $\Delta G(n)$ vary depending on the specific nucleation problem that is modelled and the level of approximation that is considered appropriate to it.

Here, we suggest that next to all the valid objections to the form of $\Delta G$ that are discussed in the literature, the structure of the Fokker-Planck equation itself must be put into question. We claim that corrections to the Fokker-Planck equation in the form of eqn.~(\ref{FP_dk}) cannot be a priori assumed to be negligible. They need to be assessed for each individual nucleation problem.

Given eqn.~(\ref{FP_dk}) we can now derive CNT as an approximation to an exact theory that has been derived from first principles (rather than to construct CNT as a phenomomenological description, as it has been done in the literature so far). The approximations that are needed to transform eqn.~(\ref{FP_dk}) into eqn.~(\ref{FP_CNT}) are the following:
\begin{itemize}
\item All coefficients $\tilde{d}_{k}(n,t)$ for $k\geq 1$ vanish.

  (Or they are such that $\int_{0}^{\infty} \tilde{d}_{k}(n,t) \mr d t = 0$ and vary on a timescale much shorter than the timescale of $p(n,t)$.)
\item The timescale of $\tilde{d}_{0}(n,t)$ is very short compared to the one of $p(n,t)$, such that we can approximate it by
  \begin{equation}
    \label{ApproxCNT1}
    \tilde{d}_{0}(n,t-\tau) = D_{0}(n)\delta(t-\tau) \end{equation}
\end{itemize}

These approximations might be appropriate in some situations, but the spectrum of processes that are referred to as nucleation phenomena is so broad that it is very unlikely that they apply in general.

Note also, that Pawula's theorem\cite{risken:1996} does not remove the discrepancies. In the Markovian case (i.e.~locality in time), Pawula's theorem would apply: the Kramers-Moyal expansion eqn.~(\ref{KM}) could then safely be truncated at order $k=2$ if at least one even coefficient $\mathcal{D}^{(2n)}$ vanished. Irrespective of whether this condition also applies here, at least the ($k=2$)-term always needs to be taken into account. This yields a term in addition to CNT, on the r.h.s. of eqn.~(\ref{KM})
\begin{align}
\frac{\partial^{2}}{\partial n^{2}} \int_{0}^{t}&\mr d\tau  \tilde{d}_{1}(n,t-\tau) \beta \frac{\partial \Delta \mathcal{G}_{\beta}^{(1)}(n) }{\partial n} p(n,\tau)
\end{align}

\section{Illustration by Molecular Dynamics Simulation}

We carried out molecular dynamics (MD) simulations of crystallization in a system of $N=32,000$ particles of mass $m$ interacting via a Lennard Jones potential
\[
V_{LJ}(r) = 4\epsilon\left(\left(\frac{\sigma}{r}\right)^{12} - \left(\frac{\sigma}{r}\right)^{6} \right)
\]
where $r$ is the distance between two particles. We used a cutoff for the potential at $2.5\; \sigma$. We simulated the dynamics in the $NVT$ ensemble with a time-step of t=0.005 $\sqrt{m\sigma^2/\epsilon}$, using a Nosé-Hoover thermostat to control the temperature. (We used this thermostat rather than a stochastic one, because the derivations presented in the previous section require deterministic microscopic dynamics.) We equilibrated the liquid phase at density $\rho\sigma^3 = 1$ and temperature $k_BT=2\epsilon$. Then we instantaneously quenched the temperature to $k_BT=0.75\epsilon$ and let the system evolve freely until it crystallized. For the chosen temperature and density, the supersaturation of the super-cooled liquid phase is moderate enough such that none of the trajectories produced more than one critical cluster within the simulated volume.

\begin{figure}
\includegraphics[width=0.99\linewidth]{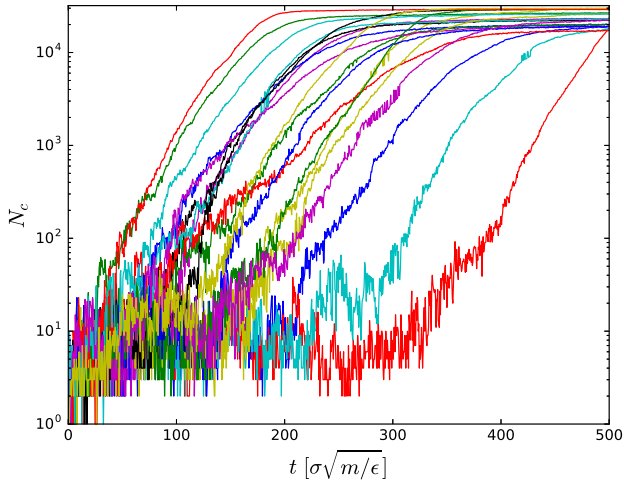}
\caption{Size of the largest crystallite as a function of time, plotted for a subset of 20 of the simulated trajectories.}
\label{f:trajectories}
\end{figure}

\begin{figure}
\centering
\includegraphics[width=0.99\linewidth]{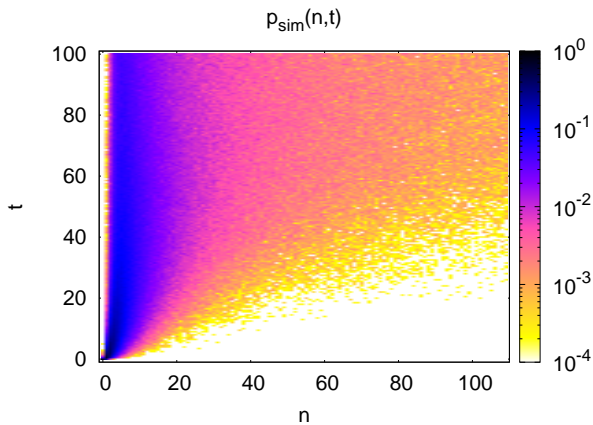}
\caption{Time-dependent distribution of the size of the largest crystallite. (Times are given in Lennard-Jones units $\sqrt{m\sigma^2/\epsilon}$.)}
\label{f:dist}
\end{figure}

\begin{figure}
\centering
\includegraphics[width=0.99\linewidth]{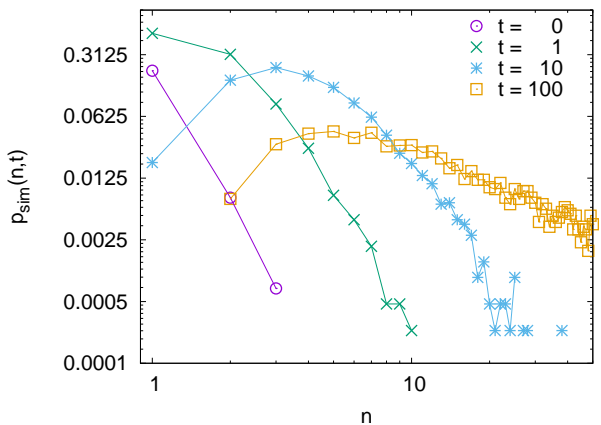}

\caption{Intersections through $p_{\rm sim}(n,t)$ for fixed times.}
\label{f:dist_intersect}
\end{figure}

In order to monitor the formation and growth of the crystallites, we used orientational bond order parameters \cite{steinhardt:1983, tenwolde:1996}. As a reaction coordinate we recorded the number $n(t)$ of particles in the largest crystalline cluster as a function of time. A total of 4262 trajectories were used for the analysis. Fig.~\ref{f:trajectories} shows the evolution of the size of the largest cluster for 20 trajectories. There is a non-zero induction time before the system nucleates and growth sets in, and the distribution of induction times is rather wide. The time-dependent distribution of cluster sizes resulting from all 4262 trajectories is shown in Fig.~\ref{f:dist} and \ref{f:dist_intersect}.

To test whether the Fokker-Planck equation is sufficient 
to describe our MD results, we compare the left-hand (lhs) and right-hand 
side (rhs) of eqn.~(\ref{FP_CNT}) for the distribution $p_{\rm sim}(n,t)$ obtained in the simulation. We computed the lhs by first smoothing $p_{\rm sim}(n,t)$
using combinations of splines and Bezier functions and then taking numerical derivatives (central differences for $t>0$; forward difference for $t=0$). The same procedure 
was applied to obtain the derivatives with respect to $n$ that appear in the rhs of eqn.~(\ref{FP_CNT}).

To construct the rhs we furthermore needed the equilibrium free energy
profile $\Delta G(n)$. We performed Monte Carlo simulations (MC) with Umbrella sampling \cite{kaestner:2011}, i.e.~rather than to employ a model for $\Delta G(n)$, which would require approximations, we determined the equilibrium cluster size distribution $p_{\beta}(n)$ by means of a separate MC simulation and then computed $\Delta G(n) = -k_BT\ln(p_{\beta}(n))$. 
We used a harmonic biasing potential on the size of the largest cluster
$w_i(n) = \dfrac{k_i}{2}\left(n-n_i\right)^2$ and overlapping windows centered
at $0\le n_i\le 109$. The strengths $k_i$ were varied such that all values
within a window were sampled; most of the windows had a width of $\Delta n=20$. \footnote{If one runs Umbrella Sampling for a large number of Monte Carlo steps, restricting the size of the largest cluster to a finite window, the system will eventually spontaneously nucleate a second large cluster in order to lower its free energy. To ensure that we sampled only one large cluster surrounded by the melt, we imposed the additional condition that the second largest cluster could not contain more than 4 particles.}

The outcome of this kind of simulation are biased probability distributions $P_i^b(n)$, which are then related to the unbiased distributions $p_{\beta}(n)$ by means of histogram reweighting.
To combine the results
of all sampling windows we used the Umbrella integration technique \cite{kaestner:2005}.
We included only those MC runs in the analysis, in which neither a trend in the average nor in
the standard deviation of the sampled cluster sizes was found. This was tested
via Mann-Kendall statistical tests with a significance level of 0.05 \cite{mann:1945}.
Finally, for $n>43$ the free energy barrier was fitted by $\partial \Delta G(n) / \partial n = f_1+f_2 n^{-1/2}$
to reduce statistical noise at high $n$ (for values $n\le43$ we used the data from the simulation directly, as the noise was negligible).

Once $\Delta G(n)$ had been determined, the only unknown term that was left on the rhs of eqn.~(\ref{FP_CNT}) was $D_0(n)$. In order to numerically determine a function $D_0(n)$ that would make the rhs equal the lhs,
we applied simulated annealing \cite{ingber:1993} to minimize
\begin{equation}
  S=\sum_{n,t} \left({\rm lhs}(n,t)-{\rm rhs}(n,t;D_0(n))\right)^2
\end{equation}

In a first attempt we assumed $D_0=$ const., i.e.~there was only one parameter to fit. We computed $S$ 
in the range $1\le n<100$ and $2.5\sqrt{m\sigma^2/\epsilon}\le t<100\sqrt{m\sigma^2/\epsilon}$ and found the best fit to be $D_0 = 0.09994$ with
$S=0.04044$. Next, we fitted $D_0(n)=D_a+D_b n^{2/3}$, i.e.~as often done in CNT, we assumed that the diffusion constant scales like the cluster surface area. With this the best fit is $D_0(n) = 0.02835 + 0.03860 n^{2/3}$, yielding
$S=0.03953$. Finally, we fitted $D_0(n)=D_a+D_b n^{D_c}$. The best fit is $D_0(n) = 0.09962 - 0.09962 n^{-3.53}$, yielding $S=0.02987$, which is still neither a particularly accurate fit, nor is there an obvious physical argument for the ${n}^{-3.53}$-dependence. In summary, using the Fokker-Planck equation we could not reproduce $p_{\rm sim}(n,t)$ well.

\begin{figure}
\centering
\subfloat{\includegraphics[width=0.49\linewidth]{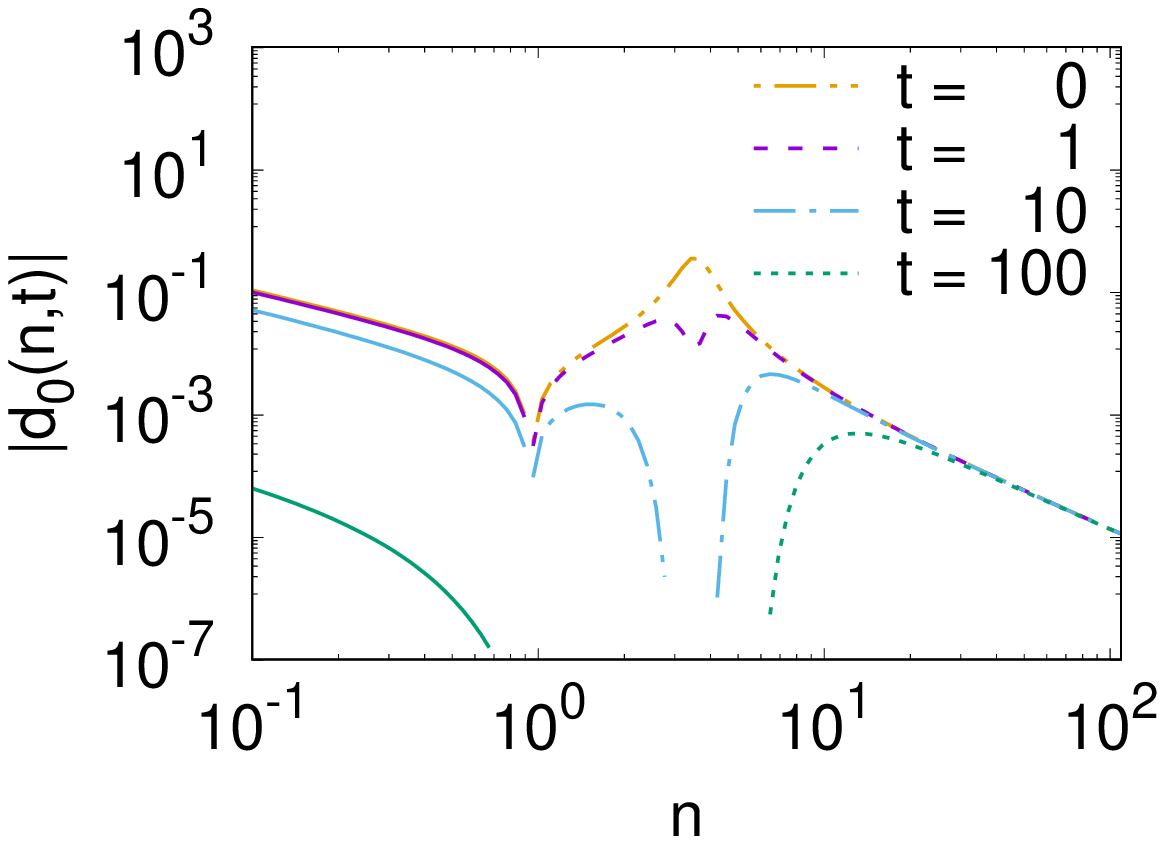}}
\subfloat{\includegraphics[width=0.49\linewidth]{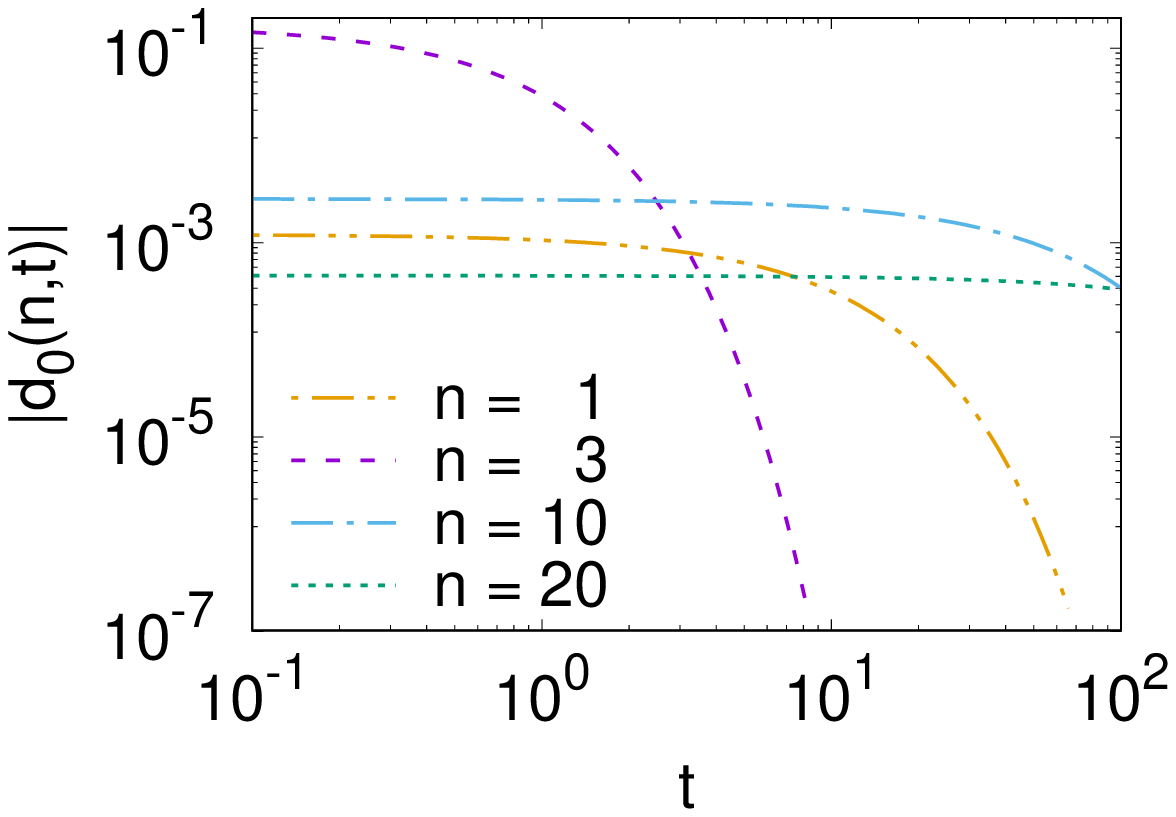}}
\caption{Intersections through the function $\tilde{d}_{0}(n,t)$, as obtained by the fit, for fixed times (left) and fixed cluster sizes (right). The time-dependence of $\tilde{d}_{0}(n,t)$ is different from a delta-distribution, thus eqn.~\ref{ApproxCNT1} is not fulfilled.}
\label{f:d0cuts}
\end{figure}

\begin{figure}
\centering
\subfloat{\includegraphics[width=0.49\linewidth]{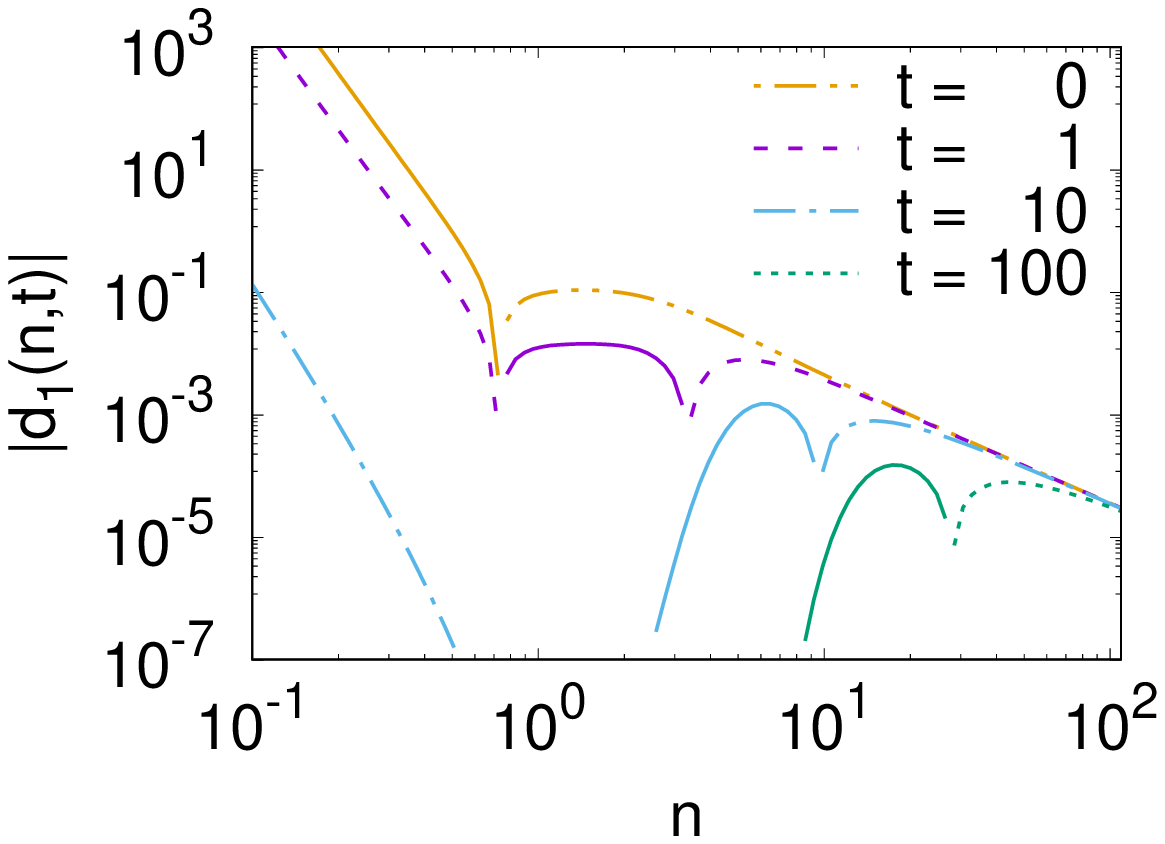}}
\subfloat{\includegraphics[width=0.49\linewidth]{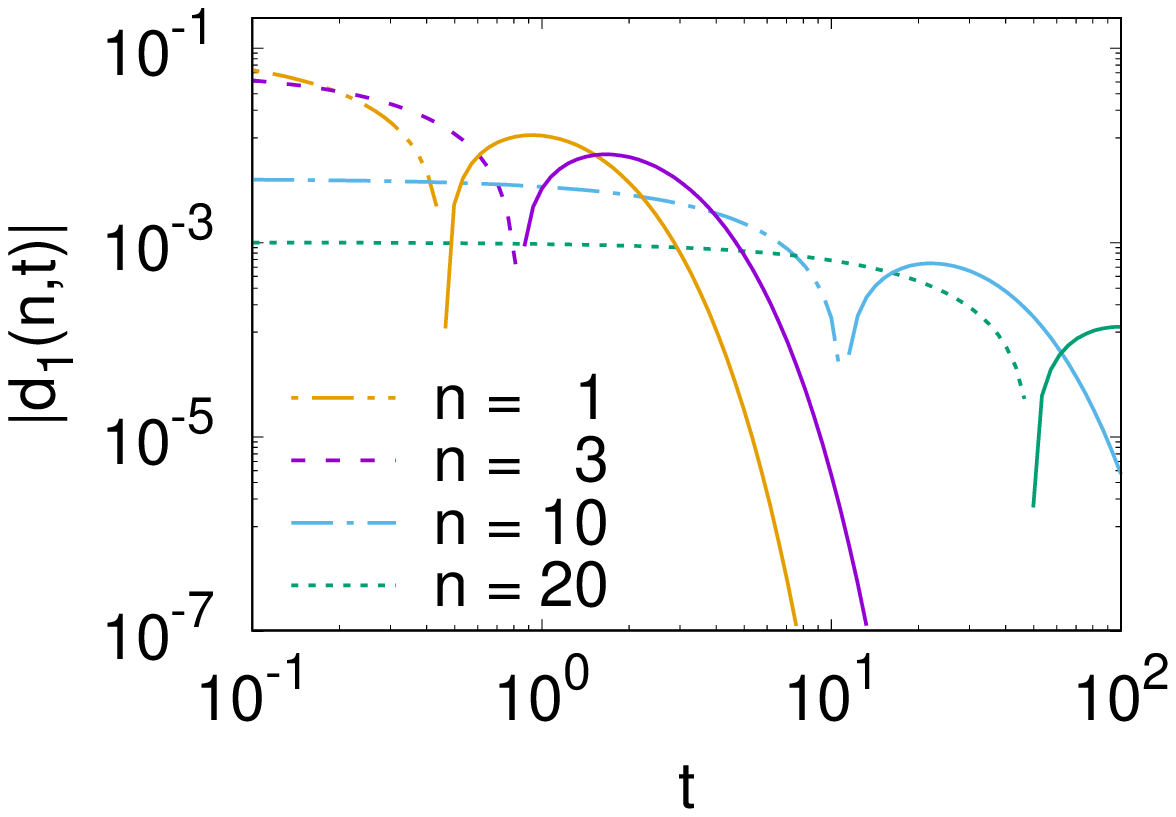}}
\caption{Intersections through $\tilde{d}_{1}(n,t)$ for fixed times (left) and fixed cluster sizes (right).}
\label{f:d1cuts}
\end{figure}

Next we applied the same strategy to eqn.~(\ref{FP_dk}). In order to limit the dimension of the parameter space for the fit, we used only the first two terms in the expansion, $\tilde{d}_0(n,t)$ and $\tilde{d}_1(n,t)$, and set the higher order terms to $0$. We made the following ansatz:

As $\tilde{d}_{0}(n,t)$ needs to become a delta-distribution in the Markovian limit, we used the form
\begin{equation}
\tilde{d}_{0}(n,t) = \frac{D_{0}(n)}{2\tau_0(n)} e^{-t/\tau_0(n)} 
\end{equation}
For $\tilde{d}_{1}(n,t)$ the Markovian limit does not necessarily require the function itself to vanish on a very short timescale, but only its integral. We therefore used
\begin{equation}
\tilde{d}_{1}(n,t) = \frac{D_1(n)}{\tau_1(n)} \left(1 - \frac{t}{\tau_1(n)} \right) 
e^{-t/\tau_1(n)}
\end{equation}
For the diffusion functions we took the same form as in the best fit of the Fokker Planck equation $D_i(n)=D_{a,i}+D_{b,i} n^{D_{c,i}}$, $i=0,1$. For the time-dependence of the kernel we used the ansatz $\tau_i = \tau_{a,i} + \tau_{b,i}n + \tau_{c,i}n^2$.
We performed simulated annealing
on $S$ as defined above, but now for eqn.~(\ref{FP_dk}). The best fit was obtained for: $\tau_0(n) = 14.2093 - 7.90387 n + 1.12207 n^2$,
$\tau_1(n) = 0.69981 - 0.37836 n + 0.14096 n^2$, 
$D_0(n) = 0.28324 - 0.26566 n^{-1.07}$, and 
$D_1(n) = 0.04872 - 0.00606 n^{-6.56}$, yielding $S=0.00975$.\footnote{It is, of course, not surprising that the quality of the fit is improved if one uses a larger number of fit parameters. However, this is not the point here. The point is that the time-scales on which the functions $\tilde{d}_{0}(n,t)$ and $\tilde{d}_{1}(n,t)$ contribute to the dynamics are significant.}
The corresponding functions $\tilde{d}_{0}(n,t)$ and $\tilde{d}_{1}(n,t)$ are shown in fig.~\ref{f:d0cuts} and \ref{f:d1cuts}. Clearly, $\tilde{d}_{0}(n,t)$ is not a delta-distribution in time. The conditions needed to obtain a Fokker-Planck equation from eqn.~(\ref{FP_dk}), are thus not fulfilled.

\begin{figure}
\centering
\subfloat{\includegraphics[width=0.99\linewidth]{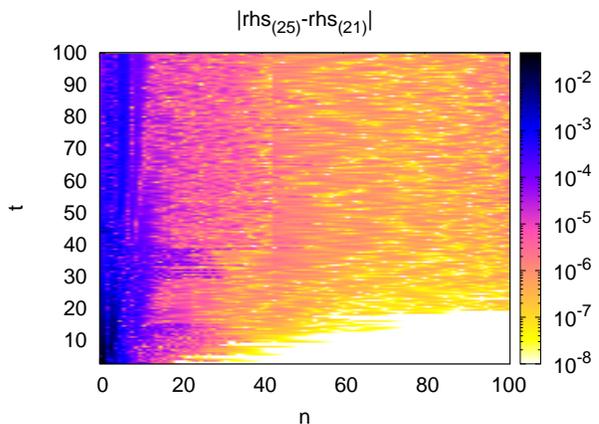}}
\caption{Difference between best fit to eqn.~(\ref{FP_CNT}) and eqn.~(\ref{FP_dk})}
\label{f:res}
\end{figure}

Fig.~(\ref{f:res}) shows the difference between the rhs of the Fokker-Planck equation, eqn.~(\ref{FP_CNT}), and eqn.~(\ref{FP_dk}) for the best fit each. There are discrepancies in particular for clusters sizes up to the top of the nucleation barrier.

The top of the free energy barrier is located at $n_{\rm crit}=33$ (see fig.~1 of the supplemental material). The corresponding mean first passage time of the MD simulation trajectories was $t\approx 80 \sqrt{m\sigma^2/\epsilon}$. 
If we compare this time to the decay time of $\tilde{d}_{0}(n,t)$, we note that the time-scale over which memory contributes to the dynamics is an order of magnitude larger than the induction time. 
We thus conclude that memory effects are relevant during crystal nucleation from the super-cooled melt and that the approximation eqn.~({\ref{ApproxCNT1}}) constitutes an over-simplification.

\section*{Conclusion}

In this paper we have presented an exact theoretical approach to nucleation based on a general, non-equilibrium, projection operator formalism. We show that Classical Nucleation Theory is a limit case of a more general theory that contains memory and out-of-equilibrium effects. In general, nucleation is neither Markovian nor diffusive. The problem can be cast in the form of a Kramers-Moyal expansion that is non-local in time and that can be related to the standard Fokker-Planck equation used in CNT. To illustrate the effect of memory, we have simulated crystallization of a supercooled Lennard-Jones melt and analyzed the cluster size distribution.

\section{Acknowldegements}
We thank T.~Franosch, Th.~Voigtmann, H.-J.~Sch\"ope, V.~Molinero, L.~Lupi, W.~Poon and G.~Cicotti for stimulating discussions. This project has been financially supported by the National Research Fund Luxembourg (FNR) within the AFR-PhD programme. Computer simulations presented in this paper were carried out using the HPC facility of the University of Luxembourg and the NEMO cluster of the University of Freiburg. We acknowledge the support by the state of Baden-W\"urttemberg through bwHPC and the German Research Foundation (DFG) through grant no INST 39/963-1 FUGG (bwFor-Cluster NEMO).

\section{Supplement material}

\subsection{Properties of the projection operator}
The projection operator $P$ is defined as
\begin{equation} 
P X(\mb \Gamma) = \int \mr d \alpha \f{\int \mr d \mb \Gamma' \rho_{\beta}(\mb \Gamma') \psi_\alpha(\mb \Gamma') X(\mb \Gamma')}{p_{\beta}(\alpha)} \psi_\alpha(\mb \Gamma) 
\end{equation}
Let us apply it to a function of the form $X(\mathbf{\Gamma}) = f(\mathbf{\Gamma})\psi_{\alpha}(\mathbf{\Gamma})$:
\begin{align} 
&P X(\mb \Gamma) = \int \mr d\alpha' \f{\int \mr d \mb \Gamma' \rho_{\beta}(\mb \Gamma') \psi_{\alpha'}(\mb \Gamma')f(\mathbf{\Gamma}')\psi_{\alpha}(\mathbf{\Gamma}')}{p_{\beta}(\alpha')} \psi_{\alpha'}(\mb \Gamma) \nonumber \\
&= \int \mr d \mb \Gamma' \rho_{\beta}(\mb \Gamma') f(\mathbf{\Gamma}') \nn \\
&\ \ \ \ \ \ \ \ \ \ \int \mr d\alpha' \f{  \delta(\alpha'-A(\mb \Gamma')) \delta(\alpha-A(\mb \Gamma'))\delta(\alpha'-A(\mb \Gamma))}{p_{\beta}(\alpha')}  \nn \\
&= \int \mr d \mb \Gamma' \rho_{\beta}(\mb \Gamma') f(\mathbf{\Gamma}') \f{  \delta(A(\mb \Gamma)-A(\mb \Gamma')) \delta(\alpha-A(\mb \Gamma'))}{p_{\beta}(A(\mb \Gamma))} \nn \\
&= \left(\int \mr d \mb \Gamma' \rho_{\beta}(\mb \Gamma') f(\mathbf{\Gamma}')   \delta(A(\mb \Gamma)-A(\mb \Gamma')) \right)\frac{ \delta(\alpha-A(\mb \Gamma))}{p_{\beta}(A(\mb \Gamma))}
\end{align}
which proves the identity $P\left[f(\mathbf{\Gamma})\psi_{\alpha}(\mathbf{\Gamma}) \right] \propto \psi_{\alpha}(\mathbf{\Gamma})$. In particular if $f(\mathbf{\Gamma}) = 1$, this relation is used to prove $P^{2}= P$.

\subsection{Detailed derivation of Grabert's formalism}

We recall here the derivation that Grabert has developed in ref.~\cite{grabert:1982} in order to obtain a Fokker-Planck-like equation for the time-dependent probability density of an arbitrary observable $A$. P being a time-independent projection operator, we start by recalling the Dyson identity:
\begin{equation*}
e^{i\mathcal{L} t} =  e^{i\mathcal{L}t}P + \int_{0}^{t}{d\tau e^{i\mathcal{L}\tau}P i\mathcal{L}  \eta(t-\tau) } + \eta(t) 
\end{equation*}
where we have defined
\begin{equation} 
\eta(t) = \left[1-P\right]e^{i\mathcal{L}t(1-P)}
\end{equation}
The Dyson decomposition can be applied on the time derivatives of the state functions $i\mathcal{L} \psi_\alpha$ reading:
\begin{align}
\label{dot_psi}
&\dot \psi_\alpha(t) =  \int \mr d \alpha' \f{\int \mr d \mb \Gamma' \rho_{\beta}(\mb \Gamma') \psi_{\alpha'}(\mb \Gamma') i\mathcal{L} \psi_{\alpha}(\mb \Gamma')}{\p_{\beta}(\alpha', t)} \psi_{\alpha'}(t) \nn \\
&+ \int_s^t \mr d\tau e^{i \mathcal{L} \tau} \nn  \\
& \ \ \ \ \ \ \ \int \mr d \alpha' \f{\int \mr d \mb \Gamma' \rho_{\beta}( \mb \Gamma') \psi_{\alpha'}(\mb \Gamma') i\mathcal{L} F_{\alpha}(t-\tau, \mb \Gamma')}{p_{\beta}(\alpha')} \psi_{\alpha'}(\tau) \nn \\
&+ e^{i\mathcal{L}s}F_{\alpha}(t-s, \mb \Gamma)
\end{align}
where we have defined
\begin{align} \label{F}
F_{\alpha}(t, \mb \Gamma) &= (1-P)e^{i\mathcal{L}t(1-P)} i\mathcal{L} \psi_\alpha (\mb \Gamma) \\
\psi_\alpha(t) &= e^{i\mathcal{L}t} \psi_\alpha(\mb \Gamma)
\end{align}
We can express the action of the Liouville operator on the state functions $\psi_{\alpha}$:
\begin{align}
i \mathcal{L} \psi_\alpha(\mb \Gamma) &= \sum_{i} \dot{\Gamma}_{i}\frac{\partial}{\partial \Gamma_{i}} \delta(A(\mb \Gamma)- \alpha) \nn \\
&= -\sum_{i} \dot{\Gamma}_{i}\frac{\partial A(\mb \Gamma)}{\partial \Gamma_{i}} \frac{\partial}{\partial\alpha} \delta(A(\mb \Gamma)- \alpha)\nn \\
& = - \f{\p}{\p \alpha} \psi_\alpha (\mb \Gamma) i \mathcal{L} A(\mb \Gamma) 
\end{align}
and therefore we can rewrite eqn.~(\ref{F}) as
\begin{equation} \label{f_alpha}
F_{\alpha}(t, \mb \Gamma) = - \f{\p}{\p \alpha} R_{\alpha}(t, \mb \Gamma)
\end{equation}
with
\begin{equation}
R_{\alpha}(t, \mb \Gamma) \equiv (1-P)e^{i\mathcal{L}t(1-P)} \psi_\alpha (\mb \Gamma) i \mathcal{L} A (\mb \Gamma)
\end{equation}
We then rewrite the first term in the r.h.s. of eqn.~(\ref{dot_psi}) as follows:
\begin{align}
&\int \mr d \alpha' \f{\int \mr d \mb \Gamma' \rho_{\beta}(\mb \Gamma') \psi_{\alpha'}(\mb \Gamma') i\mathcal{L} \psi_{\alpha}(\mb \Gamma')}{p_{\beta}(\alpha')} \psi_{\alpha'}(t)  \nn \\
&= - \f{\p}{\p \alpha} \left[ \int \mr d \alpha' \f{\int \mr d \mb \Gamma' \rho_{\beta}(\mb \Gamma')   \psi_{\alpha'}(\mb \Gamma') \psi_\alpha(\mb \Gamma') i\mathcal{L} A(\mb \Gamma') }{p_{\beta}(\alpha')}\psi_{\alpha'}(t) \right] \nn \\
&= - \f{\p}{\p \alpha} v_{\beta, \alpha}\psi_{\alpha}(t)
\end{align}
where we have defined the \textit{drift} as
\begin{equation*}
v_{\beta, \alpha} = \int \mr d \mb \Gamma \rho_{\beta}(\mb\Gamma) \psi_\alpha(\mb \Gamma) i\mathcal{L} A(\mb \Gamma)p_{\beta}(\alpha)^{-1}
\end{equation*}
To simplify the second term in the r.h.s. of eqn.~(\ref{dot_psi}) we need to introduce a new tool in the formalism: we define the transposed projector $P^T$ acting on the densities' space such that
\begin{equation} \label{p_t}
\int \mr d \mb \Gamma \mu(\mb \Gamma) P X(\mb \Gamma) = \int \mr d \mb \Gamma X (\mb \Gamma) P^T \mu (\mb \Gamma)
\end{equation}
Using the definition of $P$ as given in eqn.~(1), we can write the transposed operator as follows:
\begin{align}
&\int \mr d \mb \Gamma \mu(\mb \Gamma) P X (\mb \Gamma) \nn \\
&= \int \mr d \mb \Gamma \mu (\mb \Gamma) \int \mr d \alpha \f{\int \mr d \mb \Gamma' \rho_{\beta}(\mb \Gamma') \psi_\alpha(\mb \Gamma') X(\mb \Gamma')}{p_{\beta}(\alpha)}\psi_\alpha(\mb \Gamma)  \nn \\
&= \int \mr d \mb \Gamma'  X (\mb \Gamma') \left\{ \rho_{\beta}(\mb \Gamma') \int \mr d \alpha \f{\int \mr d \mb \Gamma \psi_\alpha (\mb \Gamma) \mu(\mb \Gamma)}{p_{\beta}(\alpha)}  \psi_\alpha(\mb \Gamma') \right\}
\end{align}
and so
\begin{equation*}
P^T \mu (\mb \Gamma) = \rho_{\beta}(\mb \Gamma) \int \mr d \alpha \f{\int \mr d \mb \Gamma' \psi_\alpha (\mb \Gamma') \mu(\mb \Gamma')}{p_{\beta}(\alpha)} \psi_\alpha(\mb \Gamma)
\end{equation*}
We can now rewrite the phase space integral in the second term of the r.h.s. of eqn.~(\ref{dot_psi}) as
\begin{align} 
\label{proto_ker}
 \int &\mr d\mb \Gamma' \rho_{\beta}( \mb \Gamma') \psi_{\alpha'}(\mb \Gamma')  i\mathcal{L}  F_{\alpha}(t-\tau, \mb \Gamma') \nn \\
 &= \int \mr d \mb\Gamma' \rho_{\beta}( \mb \Gamma') \psi_{\alpha'}(\mb \Gamma')i\mathcal{L} \left(- \f{\p}{\p \alpha} R_{\alpha}(t-\tau,\mb \Gamma') \right)  \nn \\ 
&= \f{\p}{\p \alpha} \int \mr d \mb \Gamma'  \rho_{\beta}( \mb \Gamma') R_{\alpha}(t-\tau, \mb \Gamma') i \mathcal{L} \psi_{\alpha'} (\mb \Gamma')  \nn \\
&= - \f{\p}{\p \alpha} \f{\p}{\p \alpha'}  \int  \mr d \mb \Gamma' \rho_{\beta}( \mb \Gamma') \psi_{\alpha'} (\mb \Gamma')  i\mathcal{L}  A (\mb \Gamma') R_{\alpha}(t-\tau, \mb \Gamma') 
\end{align}
where we have used eqn.~(\ref{f_alpha}); in the third identity we have used the property $\la X i\mathcal{L} Y \ra_{\beta} = -\la Y i\mathcal{L} X \ra_{\beta}$. 
This identity is only valid because $\rho_{\beta}$ is by definition a stationary distribution, which implies $i\mathcal{L}\rho_{\beta} = 0$. Thus in the standard scalar product $\left( X, Y \right) =  \la X Y \ra_{\beta}$, $i\mathcal{L}$ is anti-self-adjoint. We can now keep on working on eqn.~(\ref{proto_ker}) to turn it into a simpler form. It holds:
\begin{align}\label{ker_time_dep}
&\int  \mr d \mb \Gamma \rho_{\beta}( \mb \Gamma)  \psi_{\alpha'} (\mb \Gamma) i\mathcal{L}  A (\mb \Gamma)  R_{\alpha}(t-\tau, \mb \Gamma) \nn \\
& =\int \mr d \mb \Gamma \rho_{\beta}( \mb \Gamma) \psi_{\alpha'}(\mb \Gamma) i\mathcal{L} A(\mb\Gamma) \nn \\
&\ \ \ \ \ \ \ \ \left[ (1-P)^2 e^{i\mathcal{L}(t-\tau)(1-P)} \psi_{\alpha}(\mb \Gamma) i\mathcal{L} A(\mb \Gamma) \right] \nn \\
 & = \int \mr d \mb \Gamma \left[ (1-P)  e^{i\mathcal{L}(t-\tau)(1-P)} \psi_\alpha (\mb \Gamma) i\mathcal{L} A(\mb \Gamma) \right] \nn \\
&\ \ \ \ \ \ \ \  \ \ \  \left[   (1-P^T) \left(\rho_{\beta}( \mb \Gamma) \psi_{\alpha'}(\mb \Gamma) i\mathcal{L} A(\mb \Gamma) \right) \right]  \nn \\
& = \int \mr d \mb \Gamma \rho_{\beta}( \mb \Gamma) \left[ (1-P)  e^{i\mathcal{L}(t-\tau)(1-P)} \psi_\alpha (\mb \Gamma) i\mathcal{L} A(\mb \Gamma) \right]  \nn \\
&\ \ \ \ \ \ \ \ \ \ \ \ \ \ \ \ \ \ \ \left[    (1-P) \left(\psi_{\alpha'}(\mb \Gamma) i\mathcal{L} A(\mb \Gamma) \right) \right]  \nn \\
&= \int \mr d \mb \Gamma \rho_{\beta}( \mb \Gamma) R_{\alpha}(t-\tau, \mb \Gamma) R_{\alpha'}(0, \mb \Gamma) 
\end{align}
where in the first line we have used the property $(1-P)^2 = 1 - P$, while in the third we have used the identity $P^T(\rho_{\beta} X) = \rho_{\beta} P X$ which can be proven straightforwardly. We can now regroup everything together and rewrite eqn.~(\ref{dot_psi}) as 
\begin{widetext}
\begin{align} 
\label{dot_psi_fin}
\dot\psi_{\alpha}(t) &= - \f{\p}{\p \alpha} \left[ v_{\beta,\alpha} \psi_\alpha(t) \right] -  \int_s^t \mr d\tau e^{i \mathcal{L} \tau}  \int \mr d \alpha'  \f{ \f{\p}{\p \alpha}\f{\p}{\p \alpha'}\int  \mr d \mb \Gamma' \rho_{\beta}( \mb \Gamma')  R_{\alpha}(t-\tau, \mb \Gamma') R_{\alpha'}(0, \mb \Gamma') }{p_{\beta}(\alpha')} \psi_{\alpha'}(\tau)  + e^{i\mathcal{L}s}F_{\alpha}(t-s,\mb\Gamma) \nn \\
&= - \f{\p}{\p \alpha} \left[ v_{\beta,\alpha} \psi_\alpha(t) \right] -\f{\p}{\p \alpha}\int_s^t \mr d \tau e^{i\mathcal{L} \tau} \int \mr d \alpha'\f{\p }{\p \alpha'}\left[D(\alpha, \alpha',t-\tau)\right] \f{\psi_{\alpha'}(\tau)}{p_{\beta}(\alpha')}   + e^{i\mathcal{L}s}F_{\alpha}(t-s, \mb\Gamma) 
\end{align}
\end{widetext}
where we have defined the diffusion kernel
\begin{equation} \label{ker}
D(\alpha, \alpha', t) =  \int  \mr d \mb \Gamma' \rho_{\beta}( \mb \Gamma')  R_{\alpha}(t, \mb \Gamma') R_{\alpha'}(0, \mb \Gamma')
\end{equation}
Now, we can rewrite $p_{\beta}(\alpha)$ by using the anti-self-adjointness of $i\mathcal{L}$, i.e.
\begin{align}
	p(\alpha,t) &= \int \mr d \mb \Gamma e^{-i\mathcal{L}t}\rho(0, \mb \Gamma) \psi_\alpha(\mb \Gamma) \nn \\
	&= \int \mr d \mb \Gamma \rho(0, \mb \Gamma) e^{i\mathcal{L}t}\psi_\alpha(\mb \Gamma) = \int \mr d \mb \Gamma \rho(0, \mb \Gamma) \psi_\alpha(t)
\end{align}
Therefore, we can multiply (\ref{dot_psi_fin}) by $\rho(0,\mb\Gamma)$ and integrate over $\mathbf{\Gamma}$ to obtain eqn.~(9) in the main text. However, for this result to be exact, the average of the 'stochastic' term must vanish. In fact we have:
\begin{widetext}
\begin{align}
 \int \mr d \mb \Gamma \rho(0,\mb \Gamma) e^{i\mathcal{L}s}F_{\alpha}(t,\mb\Gamma) &= \int \mr d \mb\Gamma \rho(0,\mb\Gamma) e^{i\mathcal{L}s}(1-P) Y_{\alpha}(t,\mb \Gamma) \nn \\
&= \int \mr d \mb\Gamma \rho(s,\mb\Gamma) Y_{\alpha}(t,\mb \Gamma) - \int \mr d \mb\Gamma \rho(s,\mb\Gamma) P Y_{\alpha}(t,\mb \Gamma)\nn \\
&= \int \mr d \mb\Gamma \rho(s,\mb\Gamma) Y_{\alpha}(t,\mb \Gamma) - \int \mr d \mb\Gamma \rho(s,\mb\Gamma) \int \mr d \alpha' \f{\int \mr d\mb \Gamma' \rho_{\beta}(\mb \Gamma')\psi_{\alpha'}(\mb \Gamma') Y_{\alpha}(t,\mb \Gamma')}{p_{\beta}(\alpha')} \psi_{\alpha'}(\mb \Gamma)\nn \\
&= \int \mr d \mb\Gamma \rho(s,\mb\Gamma) Y_{\alpha}(t,\mb \Gamma) - \int \mr d \alpha'  \f{\int \mr d \mb\Gamma \rho(s,\mb\Gamma)\psi_{\alpha'}(\mb \Gamma)}{p_{\beta}(\alpha')} \int \mr d\mb \Gamma' \rho_{\beta}(\mb \Gamma')\psi_{\alpha'}(\mb \Gamma') Y_{\alpha}(t,\mb \Gamma') \nn \\
&= \int \mr d \mb\Gamma \rho(s,\mb\Gamma) Y_{\alpha}(t,\mb \Gamma) - \int \mr d \alpha'  \f{p(\alpha', s)}{p_{\beta}(\alpha')} \int \mr d\mb \Gamma \rho_{\beta}(\mb \Gamma)\psi_{\alpha'}(\mb \Gamma) Y_{\alpha}(t,\mb \Gamma) \nn \\
&= \int \mr d \mb\Gamma \rho_{\beta}(\mb\Gamma) \left[ \f{\rho(s,\mb\Gamma)}{\rho_{\beta}(\mb\Gamma)}- \int \mr d \alpha'  \f{p(\alpha',s)}{p_{\beta}(\alpha')}  \psi_{\alpha'}(\mb \Gamma) \right] Y_{\alpha}(t,\mb \Gamma) \nn \\
&= \int \mr d \mb\Gamma \rho_{\beta}(\mb\Gamma) \left[ \f{\rho(s,\mb\Gamma)}{\rho_{\beta}(\mb\Gamma)}- \f{p(A(\mb\Gamma),s)}{p_{\beta}( A(\mb \Gamma))}   \right] Y_{\alpha}(t,\mb \Gamma) 
\end{align}
\end{widetext}
where $Y_{\alpha}(t,\mathbf{\Gamma}) = e^{i\mathcal{L}(1-P)t}i\mathcal{L}\psi_{\alpha}(\mathbf{\Gamma})$. In order for the average noise to vanish, one needs the difference of the ratios in the latter equation to vanish. This is true when one works with 'relevant' variables, or relevant densities (in Grabert's meaning), i.e. the density in phase-space in fully determined by the probability density of the variable $A$.

\subsection{Expansion of the function $D(\alpha, \alpha',t)$}
We show here how we transform the non-locality in $\alpha$ in eqn.~(9), main text, into a non-Markovian Kramers-Moyal expansion. To do this, we first expand $D(\alpha,\alpha',t)$ into its Taylor series, i.e.
\begin{align}
D(\alpha,\alpha',t) = \sum_{p=0}^{\infty} \frac{t^p}{p!} \int  \mr d \mb \Gamma' \rho_{\beta}( \mb \Gamma')  R_{p}(\alpha,\mathbf{\Gamma}) R_{0}(\alpha',\mathbf{\Gamma})  
\end{align}
where we have defined
\begin{align}
R_{p}(\alpha,\mathbf{\Gamma}) &:= \left.\frac{\partial^{p} R_{\alpha}(t, \mb \Gamma')}{\partial t^{p}}\right|_{t=0} \\
&= (1-P)[i\mathcal{L}(1-P)]^{p}\left(\psi_{\alpha}(\mathbf{\Gamma})i\mathcal{L}A\mathbf{\Gamma}) \right)
\end{align}
Since we know the relation $i\mathcal{L}\psi_{\alpha} = -i\mathcal{L}A\frac{\partial \psi_{\alpha}}{\partial \alpha}$, we guess that the application of the operator $[i\mathcal{L}(1-P)]^{p}$ yields derivatives of $\psi_{\alpha}$ with respect to $\alpha$ up to order $p$. Formally, we assume
\begin{align}
\label{expand_R}
R_{p}(\alpha,\mathbf{\Gamma}) &= \sum_{k=0}^{n} r_{p,k}(\alpha,\mathbf{\Gamma})\frac{\partial^{k}\psi_{\alpha}}{\partial\alpha^{k}}
\end{align}
where the coefficients $r_{p,k}$ are defined via this equation. This identity can be proven in the following way. Assume eqn.~(\ref{expand_R}) is valid, notice that $R_{p+1} = (1-P)i\mathcal{L}R_{p}$, and thus apply the operator $(1-P)i\mathcal{L}$ to eq.~(\ref{expand_R}). The first term containing only the action of the Liouvillian is straightforwardly put into the same form as eqn.~(\ref{expand_R}), but the projected part must be taken with care. In fact, one needs to use the following relation: for any functions $f_{k}(x)$ and $g(x)$ of a vraiable $x$, one can show for any $p\in\mathbb{N}$
\begin{equation}
\label{identity_f}
\sum_{k=0}^{p} f_{k}(x)\frac{\partial^{k}g(x)}{\partial x^{p}} = \sum_{k=0}^{p} \frac{\mr d^{k}}{\mr d x^{k}} \left( h_{p,k}(x) g(x)\right) 
\end{equation}
with 
\begin{equation}
h_{p,k}(x) = \sum_{k'=k}^{p} {{k'}\choose{k}} (-1)^{k'-k} \frac{\mr d^{k'-k}f_{k'}(x)}{\mr dx^{k'-k}}
\end{equation}
This result allows to derive the following induction relation
\begin{align}
r_{p+1,0} =& i\mathcal{L} r_{p,0} - \sum_{k'=0}^{p} \frac{\partial^{k'} \left\langle \xi_{p,k'} + \chi_{p,k'} \right\rangle_{\alpha,\beta}}{\partial  \alpha^{k'}} \\
r_{p+1,k} =& i\mathcal{L} r_{p,k} - r_{p,k-1}i\mathcal{L}A \nonumber \\
&- \sum_{k'=k}^{p} {{k'}\choose{k}}  \frac{\partial^{k'-k}\left\langle \xi_{p,k'} - \chi_{p,k'} \right\rangle_{\alpha,\beta}}{\partial  \alpha^{k'-k}} \nonumber \\
& +  {{p+1}\choose{k}}  \frac{\partial^{p+1-k}\left\langle \chi_{p,p+1} \right\rangle_{\alpha,\beta}}{\partial  \alpha^{p+1-k}}
\nonumber \\
& \ \ \ \ \ \ \ \ \ \ \ \ \ \ \ \ \ \ \ \ \ \ \ \ \ \ \text{if } 0 < k < p+1 \\
r_{p+1,p+1} =& -  r_{p,p}i\mathcal{L}A + p(\chi_{p,p+1})
\end{align}
where we have defined
\begin{align}
\xi_{p,k} &=  \sum_{k'=k}^{p} {{k'}\choose{k}} (-1)^{k'-k} \frac{\mr d^{k'-k} \left( i\mathcal{L}r_{p,k'} \right)}{\mr d\alpha^{k'-k}} \\
\chi_{p,k} &=  \sum_{k'=\max\left(1,k\right)}^{n+1} {{k'}\choose{k}} (-1)^{k'-k} \frac{\mr d^{k'-k}  r_{p,k'-1} }{\mr d\alpha^{k'-k}} i\mathcal{L}A
\end{align}
The induction relation is closed by specifying the first element, namely
\begin{equation}
r_{0,k}(\alpha,\mathbf{\Gamma}) = i\mathcal{L}A(\mathbf{\Gamma}) \delta_{k,0}  
\end{equation}

Now, we use again the identity (\ref{identity_f}) to transform (\ref{expand_R}) and find
\begin{equation}
\label{expandRRn}
R_{p}(\alpha,\mathbf{\Gamma}) = \sum_{k=0}^{p} \frac{\partial^{k}}{\partial \alpha^{k}} \left( \zeta_{p,k}(\alpha,\mathbf{\Gamma}) \psi_{\alpha} \right)
\end{equation}
where we have defined 
\begin{equation}
\zeta_{p,k}(\alpha) = \sum_{k'=k}^{p} (-1)^{k'-k}{{k'}\choose{k}} \frac{\partial^{k'-k}r_{p,k'}(\alpha,\mathbf{\Gamma})} {\partial\alpha^{k'-k}}
\end{equation}
This equation is finally inserted into the Taylor series of $D(\alpha,\alpha',t)$ to find
\begin{align}
\label{expandD}
D(\alpha,\alpha',t) &=  \sum_{k=0}^{\infty}\frac{\partial^{k} }{\partial\alpha^{k}} \left[  d_{k}(\alpha,\alpha',t) \delta(\alpha-\alpha')  \right]
\end{align}
where we have defined
\begin{align}
\label{ea:defd}
&d_{k}(\alpha,\alpha',t) =  \frac{1}{p_{\beta}(\alpha')} \times \nn \\
& \int \mr d\mathbf{\Gamma} \rho_{\beta}(\mathbf{\Gamma}) \left[ \sum_{p=k}^{\infty} \zeta_{p,k}(\alpha,\mathbf{\Gamma}) \frac{t^{p}}{p!} \right] i\mathcal{L}A(\mathbf{\Gamma}) \delta(\alpha-A(\mathbf{\Gamma})) 
\end{align}
This proves the structure of eq.~(21) in the main text, and that the generalized diffusion constants $\tilde{d}_{k}(\alpha,t)=d(\alpha,\alpha',t)$ have their first $k-1$ initial time-derivatives vanishing at $t=0$.

\subsection{Odd orders of $w^{(i_{1},\cdots,i_{p})}_{\beta}(n)$}
From the definition of the functions $w^{(i_{1},\cdots,i_{p})}_{\beta}(n)$ (eqn.~(10) in the main text) we can infer the following. Since $N(\mathbf{\Gamma})$ is a function of the positions only, and thus $i\mathcal{L}N(\mathbf{\Gamma}) = \sum_{j} \frac{\text{p}_{j}}{m}\frac{\partial N}{\partial q_{j}}$, an arbitrary power of the Liouvillian $(i\mathcal{L})^{p}N$ can be written as a sum of terms, each of them being proportional to a product of momenta $\text{p}_{j_{1}}\cdots \text{p}_{j_{k}}$ such that the global power $k$ is of the same parity as $p$. The product of all these powers in $(i\mathcal{L})^{i_{1}}N\cdots(i\mathcal{L})^{i_{p}}N$, as requires the definition of $w^{(i_{1},\cdots,i_{p})}_{\beta}(n)$ can then also be decomposed into terms proportional to a power of momenta of the same parity as $\sum_{k} i_{k}$. If this quantity is odd, we thus have to average an odd power of momenta using an equilibrium measure. Since equilibrium requires an even distribution $\rho_{\beta}$ for the momenta, we conclude that $w^{(i_{1},\cdots,i_{p})}_{\beta}(n) = 0$ if $\sum_{k} i_{k}$ is odd.

\subsection{Free energy barrier}
\vspace*{-\baselineskip}
\begin{figure}[h]
\centering
\includegraphics[width=0.99\linewidth]{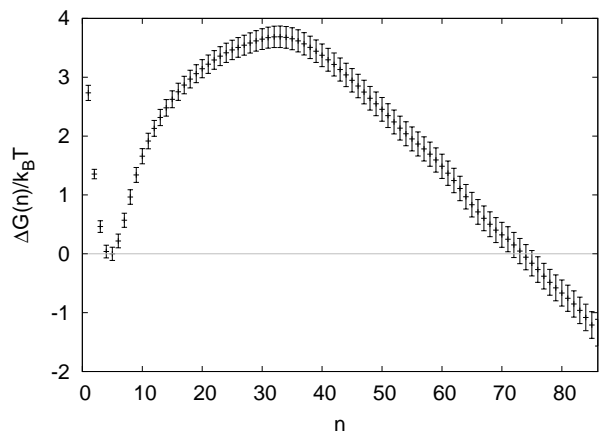}

\caption{Free energy of a cluster of size $n$ surrounded by the super-cooled melt as obtained by means of Umbrella Sampling.}
\label{f:barrier}
\end{figure}


\begin{thebibliography}{52}%
\makeatletter
\providecommand \@ifxundefined [1]{%
 \@ifx{#1\undefined}
}%
\providecommand \@ifnum [1]{%
 \ifnum #1\expandafter \@firstoftwo
 \else \expandafter \@secondoftwo
 \fi
}%
\providecommand \@ifx [1]{%
 \ifx #1\expandafter \@firstoftwo
 \else \expandafter \@secondoftwo
 \fi
}%
\providecommand \natexlab [1]{#1}%
\providecommand \enquote  [1]{``#1''}%
\providecommand \bibnamefont  [1]{#1}%
\providecommand \bibfnamefont [1]{#1}%
\providecommand \citenamefont [1]{#1}%
\providecommand \href@noop [0]{\@secondoftwo}%
\providecommand \href [0]{\begingroup \@sanitize@url \@href}%
\providecommand \@href[1]{\@@startlink{#1}\@@href}%
\providecommand \@@href[1]{\endgroup#1\@@endlink}%
\providecommand \@sanitize@url [0]{\catcode `\\12\catcode `\$12\catcode
  `\&12\catcode `\#12\catcode `\^12\catcode `\_12\catcode `\%12\relax}%
\providecommand \@@startlink[1]{}%
\providecommand \@@endlink[0]{}%
\providecommand \url  [0]{\begingroup\@sanitize@url \@url }%
\providecommand \@url [1]{\endgroup\@href {#1}{\urlprefix }}%
\providecommand \urlprefix  [0]{URL }%
\providecommand \Eprint [0]{\href }%
\providecommand \doibase [0]{http://dx.doi.org/}%
\providecommand \selectlanguage [0]{\@gobble}%
\providecommand \bibinfo  [0]{\@secondoftwo}%
\providecommand \bibfield  [0]{\@secondoftwo}%
\providecommand \translation [1]{[#1]}%
\providecommand \BibitemOpen [0]{}%
\providecommand \bibitemStop [0]{}%
\providecommand \bibitemNoStop [0]{.\EOS\space}%
\providecommand \EOS [0]{\spacefactor3000\relax}%
\providecommand \BibitemShut  [1]{\csname bibitem#1\endcsname}%
\let\auto@bib@innerbib\@empty
\bibitem [{\citenamefont {Oxtoby}(1992)}]{oxtoby:1992}%
  \BibitemOpen
  \bibfield  {author} {\bibinfo {author} {\bibfnamefont {D.~W.}\ \bibnamefont
  {Oxtoby}},\ }\href@noop {} {\bibfield  {journal} {\bibinfo  {journal}
  {Journal of Physics: Condensed Matter}\ }\textbf {\bibinfo {volume} {4}},\
  \bibinfo {pages} {7627} (\bibinfo {year} {1992})}\BibitemShut {NoStop}%
\bibitem [{\citenamefont {Kelton}\ and\ \citenamefont
  {Greer}(2010)}]{kelton:2010}%
  \BibitemOpen
  \bibfield  {author} {\bibinfo {author} {\bibfnamefont {K.}~\bibnamefont
  {Kelton}}\ and\ \bibinfo {author} {\bibfnamefont {A.~L.}\ \bibnamefont
  {Greer}},\ }\href@noop {} {\emph {\bibinfo {title} {Nucleation in condensed
  matter: applications in materials and biology}}},\ Vol.~\bibinfo {volume}
  {15}\ (\bibinfo  {publisher} {Elsevier},\ \bibinfo {year} {2010})\BibitemShut
  {NoStop}%
\bibitem [{\citenamefont {Finney}\ and\ \citenamefont
  {Finke}(2008)}]{finney:2008}%
  \BibitemOpen
  \bibfield  {author} {\bibinfo {author} {\bibfnamefont {E.~E.}\ \bibnamefont
  {Finney}}\ and\ \bibinfo {author} {\bibfnamefont {R.~G.}\ \bibnamefont
  {Finke}},\ }\href@noop {} {\bibfield  {journal} {\bibinfo  {journal} {Journal
  of Colloid and Interface Science}\ }\textbf {\bibinfo {volume} {317}},\
  \bibinfo {pages} {351} (\bibinfo {year} {2008})}\BibitemShut {NoStop}%
\bibitem [{\citenamefont {Kashchiev}(2000)}]{kashchiev:2000}%
  \BibitemOpen
  \bibfield  {author} {\bibinfo {author} {\bibfnamefont {D.}~\bibnamefont
  {Kashchiev}},\ }\href@noop {} {\emph {\bibinfo {title} {Nucleation}}}\
  (\bibinfo  {publisher} {Elsevier},\ \bibinfo {year} {2000})\BibitemShut
  {NoStop}%
\bibitem [{\citenamefont {Cantrell}\ and\ \citenamefont
  {Heymsfield}(2005)}]{cantrell:2005}%
  \BibitemOpen
  \bibfield  {author} {\bibinfo {author} {\bibfnamefont {W.}~\bibnamefont
  {Cantrell}}\ and\ \bibinfo {author} {\bibfnamefont {A.}~\bibnamefont
  {Heymsfield}},\ }\href@noop {} {\bibfield  {journal} {\bibinfo  {journal}
  {Bulletin of the American Meteorological Society}\ }\textbf {\bibinfo
  {volume} {86}},\ \bibinfo {pages} {795} (\bibinfo {year} {2005})}\BibitemShut
  {NoStop}%
\bibitem [{\citenamefont {Viisanen}\ \emph {et~al.}(1993)\citenamefont
  {Viisanen}, \citenamefont {Strey},\ and\ \citenamefont
  {Reiss}}]{viisanen:1993}%
  \BibitemOpen
  \bibfield  {author} {\bibinfo {author} {\bibfnamefont {Y.}~\bibnamefont
  {Viisanen}}, \bibinfo {author} {\bibfnamefont {R.}~\bibnamefont {Strey}}, \
  and\ \bibinfo {author} {\bibfnamefont {H.}~\bibnamefont {Reiss}},\
  }\href@noop {} {\bibfield  {journal} {\bibinfo  {journal} {The Journal of
  chemical physics}\ }\textbf {\bibinfo {volume} {99}},\ \bibinfo {pages}
  {4680} (\bibinfo {year} {1993})}\BibitemShut {NoStop}%
\bibitem [{\citenamefont {Zhang}\ \emph {et~al.}(2011)\citenamefont {Zhang},
  \citenamefont {Khalizov}, \citenamefont {Wang}, \citenamefont {Hu},\ and\
  \citenamefont {Xu}}]{zhang:2011}%
  \BibitemOpen
  \bibfield  {author} {\bibinfo {author} {\bibfnamefont {R.}~\bibnamefont
  {Zhang}}, \bibinfo {author} {\bibfnamefont {A.}~\bibnamefont {Khalizov}},
  \bibinfo {author} {\bibfnamefont {L.}~\bibnamefont {Wang}}, \bibinfo {author}
  {\bibfnamefont {M.}~\bibnamefont {Hu}}, \ and\ \bibinfo {author}
  {\bibfnamefont {W.}~\bibnamefont {Xu}},\ }\href@noop {} {\bibfield  {journal}
  {\bibinfo  {journal} {Chemical Reviews}\ }\textbf {\bibinfo {volume} {112}},\
  \bibinfo {pages} {1957} (\bibinfo {year} {2011})}\BibitemShut {NoStop}%
\bibitem [{\citenamefont {Volmer}\ and\ \citenamefont
  {Weber}(1926)}]{volmer:1926}%
  \BibitemOpen
  \bibfield  {author} {\bibinfo {author} {\bibfnamefont {M.}~\bibnamefont
  {Volmer}}\ and\ \bibinfo {author} {\bibfnamefont {A.}~\bibnamefont {Weber}},\
  }\href@noop {} {\bibfield  {journal} {\bibinfo  {journal} {Zeitschrift
  f{\"u}r physikalische Chemie}\ }\textbf {\bibinfo {volume} {119}},\ \bibinfo
  {pages} {277} (\bibinfo {year} {1926})}\BibitemShut {NoStop}%
\bibitem [{\citenamefont {Becker}\ and\ \citenamefont
  {D{\"o}ring}(1935)}]{becker:1935}%
  \BibitemOpen
  \bibfield  {author} {\bibinfo {author} {\bibfnamefont {R.}~\bibnamefont
  {Becker}}\ and\ \bibinfo {author} {\bibfnamefont {W.}~\bibnamefont
  {D{\"o}ring}},\ }\href@noop {} {\bibfield  {journal} {\bibinfo  {journal}
  {Annalen der Physik}\ }\textbf {\bibinfo {volume} {416}},\ \bibinfo {pages}
  {719} (\bibinfo {year} {1935})}\BibitemShut {NoStop}%
\bibitem [{\citenamefont {Kalikmanov}(2013)}]{kalikmanov:2013}%
  \BibitemOpen
  \bibfield  {author} {\bibinfo {author} {\bibfnamefont {V.~I.}\ \bibnamefont
  {Kalikmanov}},\ }in\ \href@noop {} {\emph {\bibinfo {booktitle} {Nucleation
  theory}}}\ (\bibinfo  {publisher} {Springer},\ \bibinfo {year} {2013})\ pp.\
  \bibinfo {pages} {17--41}\BibitemShut {NoStop}%
\bibitem [{\citenamefont {Ko{\v{z}}{\'\i}{\v{s}}ek}\ and\ \citenamefont
  {Demo}(2017)}]{kovzivsek:2017}%
  \BibitemOpen
  \bibfield  {author} {\bibinfo {author} {\bibfnamefont {Z.}~\bibnamefont
  {Ko{\v{z}}{\'\i}{\v{s}}ek}}\ and\ \bibinfo {author} {\bibfnamefont
  {P.}~\bibnamefont {Demo}},\ }\href@noop {} {\bibfield  {journal} {\bibinfo
  {journal} {Journal of Crystal Growth}\ }\textbf {\bibinfo {volume} {475}},\
  \bibinfo {pages} {247} (\bibinfo {year} {2017})}\BibitemShut {NoStop}%
\bibitem [{\citenamefont {Ayuba}\ \emph {et~al.}(2018)\citenamefont {Ayuba},
  \citenamefont {Suh}, \citenamefont {Nomura}, \citenamefont {Ebisuzaki},\ and\
  \citenamefont {Yasuoka}}]{ayuba:2018}%
  \BibitemOpen
  \bibfield  {author} {\bibinfo {author} {\bibfnamefont {S.}~\bibnamefont
  {Ayuba}}, \bibinfo {author} {\bibfnamefont {D.}~\bibnamefont {Suh}}, \bibinfo
  {author} {\bibfnamefont {K.}~\bibnamefont {Nomura}}, \bibinfo {author}
  {\bibfnamefont {T.}~\bibnamefont {Ebisuzaki}}, \ and\ \bibinfo {author}
  {\bibfnamefont {K.}~\bibnamefont {Yasuoka}},\ }\href@noop {} {\bibfield
  {journal} {\bibinfo  {journal} {The Journal of chemical physics}\ }\textbf
  {\bibinfo {volume} {149}},\ \bibinfo {pages} {044504} (\bibinfo {year}
  {2018})}\BibitemShut {NoStop}%
\bibitem [{\citenamefont {Richard}\ and\ \citenamefont
  {Speck}(2018)}]{richard:2018}%
  \BibitemOpen
  \bibfield  {author} {\bibinfo {author} {\bibfnamefont {D.}~\bibnamefont
  {Richard}}\ and\ \bibinfo {author} {\bibfnamefont {T.}~\bibnamefont
  {Speck}},\ }\href@noop {} {\bibfield  {journal} {\bibinfo  {journal} {The
  Journal of chemical physics}\ }\textbf {\bibinfo {volume} {148}},\ \bibinfo
  {pages} {224102} (\bibinfo {year} {2018})}\BibitemShut {NoStop}%
\bibitem [{\citenamefont {Dumitrescu}\ \emph {et~al.}(2017)\citenamefont
  {Dumitrescu}, \citenamefont {Smeulders}, \citenamefont {Dam},\ and\
  \citenamefont {Gaastra-Nedea}}]{dumitrescu:2017}%
  \BibitemOpen
  \bibfield  {author} {\bibinfo {author} {\bibfnamefont {L.~R.}\ \bibnamefont
  {Dumitrescu}}, \bibinfo {author} {\bibfnamefont {D.~M.}\ \bibnamefont
  {Smeulders}}, \bibinfo {author} {\bibfnamefont {J.~A.}\ \bibnamefont {Dam}},
  \ and\ \bibinfo {author} {\bibfnamefont {S.~V.}\ \bibnamefont
  {Gaastra-Nedea}},\ }\href@noop {} {\bibfield  {journal} {\bibinfo  {journal}
  {The Journal of Chemical Physics}\ }\textbf {\bibinfo {volume} {146}},\
  \bibinfo {pages} {084309} (\bibinfo {year} {2017})}\BibitemShut {NoStop}%
\bibitem [{\citenamefont {Marchio}\ \emph {et~al.}(2018)\citenamefont
  {Marchio}, \citenamefont {Meloni}, \citenamefont {Giacomello}, \citenamefont
  {Valeriani},\ and\ \citenamefont {Casciola}}]{marchio:2018}%
  \BibitemOpen
  \bibfield  {author} {\bibinfo {author} {\bibfnamefont {S.}~\bibnamefont
  {Marchio}}, \bibinfo {author} {\bibfnamefont {S.}~\bibnamefont {Meloni}},
  \bibinfo {author} {\bibfnamefont {A.}~\bibnamefont {Giacomello}}, \bibinfo
  {author} {\bibfnamefont {C.}~\bibnamefont {Valeriani}}, \ and\ \bibinfo
  {author} {\bibfnamefont {C.~M.}\ \bibnamefont {Casciola}},\ }\href {\doibase
  10.1063/1.5011106} {\bibfield  {journal} {\bibinfo  {journal} {JOURNAL OF
  CHEMICAL PHYSICS}\ }\textbf {\bibinfo {volume} {148}} (\bibinfo {year}
  {2018}),\ 10.1063/1.5011106}\BibitemShut {NoStop}%
\bibitem [{\citenamefont {Leines}\ \emph {et~al.}(2017)\citenamefont {Leines},
  \citenamefont {Drautz},\ and\ \citenamefont {Rogal}}]{leines:2017}%
  \BibitemOpen
  \bibfield  {author} {\bibinfo {author} {\bibfnamefont {G.~D.}\ \bibnamefont
  {Leines}}, \bibinfo {author} {\bibfnamefont {R.}~\bibnamefont {Drautz}}, \
  and\ \bibinfo {author} {\bibfnamefont {J.}~\bibnamefont {Rogal}},\ }\href
  {\doibase 10.1063/1.4980082} {\bibfield  {journal} {\bibinfo  {journal}
  {JOURNAL OF CHEMICAL PHYSICS}\ }\textbf {\bibinfo {volume} {146}} (\bibinfo
  {year} {2017}),\ 10.1063/1.4980082}\BibitemShut {NoStop}%
\bibitem [{\citenamefont {Desgranges}\ and\ \citenamefont
  {Delhommelle}(2018)}]{desgranges:2018}%
  \BibitemOpen
  \bibfield  {author} {\bibinfo {author} {\bibfnamefont {C.}~\bibnamefont
  {Desgranges}}\ and\ \bibinfo {author} {\bibfnamefont {J.}~\bibnamefont
  {Delhommelle}},\ }\href {\doibase 10.1103/PhysRevLett.120.115701} {\bibfield
  {journal} {\bibinfo  {journal} {PHYSICAL REVIEW LETTERS}\ }\textbf {\bibinfo
  {volume} {120}} (\bibinfo {year} {2018}),\
  10.1103/PhysRevLett.120.115701}\BibitemShut {NoStop}%
\bibitem [{\citenamefont {Jiang}\ \emph {et~al.}(2018)\citenamefont {Jiang},
  \citenamefont {Haji-Akbari}, \citenamefont {Debenedetti},\ and\ \citenamefont
  {Panagiotopoulos}}]{jiang:2018}%
  \BibitemOpen
  \bibfield  {author} {\bibinfo {author} {\bibfnamefont {H.}~\bibnamefont
  {Jiang}}, \bibinfo {author} {\bibfnamefont {A.}~\bibnamefont {Haji-Akbari}},
  \bibinfo {author} {\bibfnamefont {P.~G.}\ \bibnamefont {Debenedetti}}, \ and\
  \bibinfo {author} {\bibfnamefont {A.~Z.}\ \bibnamefont {Panagiotopoulos}},\
  }\href {\doibase 10.1063/1.5016554} {\bibfield  {journal} {\bibinfo
  {journal} {JOURNAL OF CHEMICAL PHYSICS}\ }\textbf {\bibinfo {volume} {148}}
  (\bibinfo {year} {2018}),\ 10.1063/1.5016554}\BibitemShut {NoStop}%
\bibitem [{\citenamefont {Winkelmann}\ \emph {et~al.}(2018)\citenamefont
  {Winkelmann}, \citenamefont {Kuczaj}, \citenamefont {Nordlund},\ and\
  \citenamefont {Geurts}}]{winkelmann:2018}%
  \BibitemOpen
  \bibfield  {author} {\bibinfo {author} {\bibfnamefont {C.}~\bibnamefont
  {Winkelmann}}, \bibinfo {author} {\bibfnamefont {A.~K.}\ \bibnamefont
  {Kuczaj}}, \bibinfo {author} {\bibfnamefont {M.}~\bibnamefont {Nordlund}}, \
  and\ \bibinfo {author} {\bibfnamefont {B.~J.}\ \bibnamefont {Geurts}},\
  }\href {\doibase 10.1007/s10665-017-9918-6} {\bibfield  {journal} {\bibinfo
  {journal} {Journal of Engineering Mathematics}\ }\textbf {\bibinfo {volume}
  {108}},\ \bibinfo {pages} {171} (\bibinfo {year} {2018})}\BibitemShut
  {NoStop}%
\bibitem [{\citenamefont {Gebauer}\ and\ \citenamefont
  {C{\"o}lfen}(2011)}]{gebauer:2011}%
  \BibitemOpen
  \bibfield  {author} {\bibinfo {author} {\bibfnamefont {D.}~\bibnamefont
  {Gebauer}}\ and\ \bibinfo {author} {\bibfnamefont {H.}~\bibnamefont
  {C{\"o}lfen}},\ }\href@noop {} {\bibfield  {journal} {\bibinfo  {journal}
  {Nano Today}\ }\textbf {\bibinfo {volume} {6}},\ \bibinfo {pages} {564}
  (\bibinfo {year} {2011})}\BibitemShut {NoStop}%
\bibitem [{\citenamefont {Hegg}\ and\ \citenamefont {Baker}(2009)}]{hegg:2009}%
  \BibitemOpen
  \bibfield  {author} {\bibinfo {author} {\bibfnamefont {D.}~\bibnamefont
  {Hegg}}\ and\ \bibinfo {author} {\bibfnamefont {M.}~\bibnamefont {Baker}},\
  }\href@noop {} {\bibfield  {journal} {\bibinfo  {journal} {Reports on
  progress in Physics}\ }\textbf {\bibinfo {volume} {72}},\ \bibinfo {pages}
  {056801} (\bibinfo {year} {2009})}\BibitemShut {NoStop}%
\bibitem [{\citenamefont {Sear}(2012)}]{sear:2012}%
  \BibitemOpen
  \bibfield  {author} {\bibinfo {author} {\bibfnamefont {R.~P.}\ \bibnamefont
  {Sear}},\ }\href@noop {} {\bibfield  {journal} {\bibinfo  {journal}
  {International Materials Reviews}\ }\textbf {\bibinfo {volume} {57}},\
  \bibinfo {pages} {328} (\bibinfo {year} {2012})}\BibitemShut {NoStop}%
\bibitem [{\citenamefont {Filion}\ \emph {et~al.}(2010)\citenamefont {Filion},
  \citenamefont {Hermes}, \citenamefont {Ni},\ and\ \citenamefont
  {Dijkstra}}]{filion:2010}%
  \BibitemOpen
  \bibfield  {author} {\bibinfo {author} {\bibfnamefont {L.}~\bibnamefont
  {Filion}}, \bibinfo {author} {\bibfnamefont {M.}~\bibnamefont {Hermes}},
  \bibinfo {author} {\bibfnamefont {R.}~\bibnamefont {Ni}}, \ and\ \bibinfo
  {author} {\bibfnamefont {M.}~\bibnamefont {Dijkstra}},\ }\href@noop {}
  {\bibfield  {journal} {\bibinfo  {journal} {The Journal of chemical physics}\
  }\textbf {\bibinfo {volume} {133}},\ \bibinfo {pages} {244115} (\bibinfo
  {year} {2010})}\BibitemShut {NoStop}%
\bibitem [{\citenamefont {Horsch}\ \emph {et~al.}(2008)\citenamefont {Horsch},
  \citenamefont {Vrabec},\ and\ \citenamefont {Hasse}}]{horsch:2008}%
  \BibitemOpen
  \bibfield  {author} {\bibinfo {author} {\bibfnamefont {M.}~\bibnamefont
  {Horsch}}, \bibinfo {author} {\bibfnamefont {J.}~\bibnamefont {Vrabec}}, \
  and\ \bibinfo {author} {\bibfnamefont {H.}~\bibnamefont {Hasse}},\ }\href
  {\doibase 10.1103/PhysRevE.78.011603} {\bibfield  {journal} {\bibinfo
  {journal} {Phys. Rev. E}\ }\textbf {\bibinfo {volume} {78}},\ \bibinfo
  {pages} {011603} (\bibinfo {year} {2008})}\BibitemShut {NoStop}%
\bibitem [{\citenamefont {Tanaka}\ \emph {et~al.}(2005)\citenamefont {Tanaka},
  \citenamefont {Kawamura}, \citenamefont {Tanaka},\ and\ \citenamefont
  {Nakazawa}}]{tanaka:2005}%
  \BibitemOpen
  \bibfield  {author} {\bibinfo {author} {\bibfnamefont {K.~K.}\ \bibnamefont
  {Tanaka}}, \bibinfo {author} {\bibfnamefont {K.}~\bibnamefont {Kawamura}},
  \bibinfo {author} {\bibfnamefont {H.}~\bibnamefont {Tanaka}}, \ and\ \bibinfo
  {author} {\bibfnamefont {K.}~\bibnamefont {Nakazawa}},\ }\href@noop {}
  {\bibfield  {journal} {\bibinfo  {journal} {The Journal of chemical physics}\
  }\textbf {\bibinfo {volume} {122}},\ \bibinfo {pages} {184514} (\bibinfo
  {year} {2005})}\BibitemShut {NoStop}%
\bibitem [{\citenamefont {Russo}\ and\ \citenamefont
  {Tanaka}(2012)}]{russo:2012}%
  \BibitemOpen
  \bibfield  {author} {\bibinfo {author} {\bibfnamefont {J.}~\bibnamefont
  {Russo}}\ and\ \bibinfo {author} {\bibfnamefont {H.}~\bibnamefont {Tanaka}},\
  }\href@noop {} {\bibfield  {journal} {\bibinfo  {journal} {Scientific
  reports}\ }\textbf {\bibinfo {volume} {2}},\ \bibinfo {pages} {505} (\bibinfo
  {year} {2012})}\BibitemShut {NoStop}%
\bibitem [{\citenamefont {Grabert}(1982)}]{grabert:1982}%
  \BibitemOpen
  \bibfield  {author} {\bibinfo {author} {\bibfnamefont {H.}~\bibnamefont
  {Grabert}},\ }\href@noop {} {\emph {\bibinfo {title} {Projection operator
  techniques in nonequilibrium statistical mechanics}}},\ Vol.~\bibinfo
  {volume} {95}\ (\bibinfo  {publisher} {Springer},\ \bibinfo {year}
  {1982})\BibitemShut {NoStop}%
\bibitem [{\citenamefont {Hansen}\ and\ \citenamefont
  {McDonald}(1990)}]{hansen:1990}%
  \BibitemOpen
  \bibfield  {author} {\bibinfo {author} {\bibfnamefont {J.-P.}\ \bibnamefont
  {Hansen}}\ and\ \bibinfo {author} {\bibfnamefont {I.~R.}\ \bibnamefont
  {McDonald}},\ }\href@noop {} {\emph {\bibinfo {title} {Theory of simple
  liquids}}}\ (\bibinfo  {publisher} {Elsevier},\ \bibinfo {year}
  {1990})\BibitemShut {NoStop}%
\bibitem [{\citenamefont {Risken}(1996)}]{risken:1996}%
  \BibitemOpen
  \bibfield  {author} {\bibinfo {author} {\bibfnamefont {H.}~\bibnamefont
  {Risken}},\ }in\ \href@noop {} {\emph {\bibinfo {booktitle} {The
  Fokker-Planck Equation}}}\ (\bibinfo  {publisher} {Springer},\ \bibinfo
  {year} {1996})\ pp.\ \bibinfo {pages} {63--95}\BibitemShut {NoStop}%
\bibitem [{\citenamefont {Laaksonen}\ \emph {et~al.}(1995)\citenamefont
  {Laaksonen}, \citenamefont {Talanquer},\ and\ \citenamefont
  {Oxtoby}}]{laaksonen:1995}%
  \BibitemOpen
  \bibfield  {author} {\bibinfo {author} {\bibfnamefont {A.}~\bibnamefont
  {Laaksonen}}, \bibinfo {author} {\bibfnamefont {V.}~\bibnamefont
  {Talanquer}}, \ and\ \bibinfo {author} {\bibfnamefont {D.~W.}\ \bibnamefont
  {Oxtoby}},\ }\href {\doibase 10.1146/annurev.pc.46.100195.002421} {\bibfield
  {journal} {\bibinfo  {journal} {Annual Review of Physical Chemistry}\
  }\textbf {\bibinfo {volume} {46}},\ \bibinfo {pages} {489} (\bibinfo {year}
  {1995})}\BibitemShut {NoStop}%
\bibitem [{\citenamefont {Ford}(2004)}]{ford:2004}%
  \BibitemOpen
  \bibfield  {author} {\bibinfo {author} {\bibfnamefont {I.~J.}\ \bibnamefont
  {Ford}},\ }\href {\doibase 10.1243/0954406041474183} {\bibfield  {journal}
  {\bibinfo  {journal} {Proceedings of the Institution of Mechanical Engineers,
  Part C: Journal of Mechanical Engineering Science}\ }\textbf {\bibinfo
  {volume} {218}},\ \bibinfo {pages} {883} (\bibinfo {year}
  {2004})}\BibitemShut {NoStop}%
\bibitem [{\citenamefont {Moroni}\ \emph {et~al.}(2005)\citenamefont {Moroni},
  \citenamefont {Ten~Wolde},\ and\ \citenamefont {Bolhuis}}]{moroni:2005}%
  \BibitemOpen
  \bibfield  {author} {\bibinfo {author} {\bibfnamefont {D.}~\bibnamefont
  {Moroni}}, \bibinfo {author} {\bibfnamefont {P.~R.}\ \bibnamefont
  {Ten~Wolde}}, \ and\ \bibinfo {author} {\bibfnamefont {P.~G.}\ \bibnamefont
  {Bolhuis}},\ }\href@noop {} {\bibfield  {journal} {\bibinfo  {journal}
  {Physical review letters}\ }\textbf {\bibinfo {volume} {94}},\ \bibinfo
  {pages} {235703} (\bibinfo {year} {2005})}\BibitemShut {NoStop}%
\bibitem [{\citenamefont {Peters}\ and\ \citenamefont
  {Trout}(2006)}]{peters:2006}%
  \BibitemOpen
  \bibfield  {author} {\bibinfo {author} {\bibfnamefont {B.}~\bibnamefont
  {Peters}}\ and\ \bibinfo {author} {\bibfnamefont {B.~L.}\ \bibnamefont
  {Trout}},\ }\href@noop {} {\bibfield  {journal} {\bibinfo  {journal} {The
  Journal of chemical physics}\ }\textbf {\bibinfo {volume} {125}},\ \bibinfo
  {pages} {054108} (\bibinfo {year} {2006})}\BibitemShut {NoStop}%
\bibitem [{\citenamefont {Barnes}\ \emph {et~al.}(2014)\citenamefont {Barnes},
  \citenamefont {Knott}, \citenamefont {Beckham}, \citenamefont {Wu},\ and\
  \citenamefont {Sum}}]{barnes:2014}%
  \BibitemOpen
  \bibfield  {author} {\bibinfo {author} {\bibfnamefont {B.~C.}\ \bibnamefont
  {Barnes}}, \bibinfo {author} {\bibfnamefont {B.~C.}\ \bibnamefont {Knott}},
  \bibinfo {author} {\bibfnamefont {G.~T.}\ \bibnamefont {Beckham}}, \bibinfo
  {author} {\bibfnamefont {D.~T.}\ \bibnamefont {Wu}}, \ and\ \bibinfo {author}
  {\bibfnamefont {A.~K.}\ \bibnamefont {Sum}},\ }\href@noop {} {\bibfield
  {journal} {\bibinfo  {journal} {The Journal of Physical Chemistry B}\
  }\textbf {\bibinfo {volume} {118}},\ \bibinfo {pages} {13236} (\bibinfo
  {year} {2014})}\BibitemShut {NoStop}%
\bibitem [{\citenamefont {Lechner}\ \emph {et~al.}(2011)\citenamefont
  {Lechner}, \citenamefont {Dellago},\ and\ \citenamefont
  {Bolhuis}}]{lechner:2011}%
  \BibitemOpen
  \bibfield  {author} {\bibinfo {author} {\bibfnamefont {W.}~\bibnamefont
  {Lechner}}, \bibinfo {author} {\bibfnamefont {C.}~\bibnamefont {Dellago}}, \
  and\ \bibinfo {author} {\bibfnamefont {P.~G.}\ \bibnamefont {Bolhuis}},\
  }\href@noop {} {\bibfield  {journal} {\bibinfo  {journal} {The Journal of
  chemical physics}\ }\textbf {\bibinfo {volume} {135}},\ \bibinfo {pages}
  {154110} (\bibinfo {year} {2011})}\BibitemShut {NoStop}%
\bibitem [{\citenamefont {{Schweitzer}}\ \emph {et~al.}(1988)\citenamefont
  {{Schweitzer}}, \citenamefont {{Schimansky-Geier}}, \citenamefont
  {{Ebeling}},\ and\ \citenamefont {{Ulbricht}}}]{schweitzer:1988}%
  \BibitemOpen
  \bibfield  {author} {\bibinfo {author} {\bibfnamefont {F.}~\bibnamefont
  {{Schweitzer}}}, \bibinfo {author} {\bibfnamefont {L.}~\bibnamefont
  {{Schimansky-Geier}}}, \bibinfo {author} {\bibfnamefont {W.}~\bibnamefont
  {{Ebeling}}}, \ and\ \bibinfo {author} {\bibfnamefont {H.}~\bibnamefont
  {{Ulbricht}}},\ }\href {\doibase 10.1016/0378-4371(88)90059-3} {\bibfield
  {journal} {\bibinfo  {journal} {Physica A Statistical Mechanics and its
  Applications}\ }\textbf {\bibinfo {volume} {150}},\ \bibinfo {pages} {261}
  (\bibinfo {year} {1988})}\BibitemShut {NoStop}%
\bibitem [{\citenamefont {Ford}(1997)}]{ford:1997}%
  \BibitemOpen
  \bibfield  {author} {\bibinfo {author} {\bibfnamefont {I.}~\bibnamefont
  {Ford}},\ }\href@noop {} {\bibfield  {journal} {\bibinfo  {journal} {Physical
  Review E}\ }\textbf {\bibinfo {volume} {56}},\ \bibinfo {pages} {5615}
  (\bibinfo {year} {1997})}\BibitemShut {NoStop}%
\bibitem [{\citenamefont {O’Malley}\ and\ \citenamefont
  {Snook}(2003)}]{omalley:2003}%
  \BibitemOpen
  \bibfield  {author} {\bibinfo {author} {\bibfnamefont {B.}~\bibnamefont
  {O’Malley}}\ and\ \bibinfo {author} {\bibfnamefont {I.}~\bibnamefont
  {Snook}},\ }\href@noop {} {\bibfield  {journal} {\bibinfo  {journal}
  {Physical review letters}\ }\textbf {\bibinfo {volume} {90}},\ \bibinfo
  {pages} {085702} (\bibinfo {year} {2003})}\BibitemShut {NoStop}%
\bibitem [{\citenamefont {Sanz}\ \emph {et~al.}(2007)\citenamefont {Sanz},
  \citenamefont {Valeriani}, \citenamefont {Frenkel},\ and\ \citenamefont
  {Dijkstra}}]{sanz:2007}%
  \BibitemOpen
  \bibfield  {author} {\bibinfo {author} {\bibfnamefont {E.}~\bibnamefont
  {Sanz}}, \bibinfo {author} {\bibfnamefont {C.}~\bibnamefont {Valeriani}},
  \bibinfo {author} {\bibfnamefont {D.}~\bibnamefont {Frenkel}}, \ and\
  \bibinfo {author} {\bibfnamefont {M.}~\bibnamefont {Dijkstra}},\ }\href@noop
  {} {\bibfield  {journal} {\bibinfo  {journal} {Physical review letters}\
  }\textbf {\bibinfo {volume} {99}},\ \bibinfo {pages} {055501} (\bibinfo
  {year} {2007})}\BibitemShut {NoStop}%
\bibitem [{\citenamefont {Binder}\ and\ \citenamefont
  {Virnau}(2016)}]{binder:2016}%
  \BibitemOpen
  \bibfield  {author} {\bibinfo {author} {\bibfnamefont {K.}~\bibnamefont
  {Binder}}\ and\ \bibinfo {author} {\bibfnamefont {P.}~\bibnamefont
  {Virnau}},\ }\href {\doibase 10.1063/1.4959235} {\bibfield  {journal}
  {\bibinfo  {journal} {The Journal of Chemical Physics}\ }\textbf {\bibinfo
  {volume} {145}},\ \bibinfo {pages} {211701} (\bibinfo {year}
  {2016})}\BibitemShut {NoStop}%
\bibitem [{\citenamefont {Oxtoby}\ and\ \citenamefont
  {Evans}(1988)}]{oxtoby:1988}%
  \BibitemOpen
  \bibfield  {author} {\bibinfo {author} {\bibfnamefont {D.~W.}\ \bibnamefont
  {Oxtoby}}\ and\ \bibinfo {author} {\bibfnamefont {R.}~\bibnamefont {Evans}},\
  }\href {\doibase 10.1063/1.455285} {\bibfield  {journal} {\bibinfo  {journal}
  {The Journal of Chemical Physics}\ }\textbf {\bibinfo {volume} {89}},\
  \bibinfo {pages} {7521} (\bibinfo {year} {1988})}\BibitemShut {NoStop}%
\bibitem [{\citenamefont {Oxtoby}(1998)}]{oxtoby:1998}%
  \BibitemOpen
  \bibfield  {author} {\bibinfo {author} {\bibfnamefont {D.}~\bibnamefont
  {Oxtoby}},\ }\href {\doibase 10.1021/ar9702278} {\bibfield  {journal}
  {\bibinfo  {journal} {ACCOUNTS OF CHEMICAL RESEARCH}\ }\textbf {\bibinfo
  {volume} {31}},\ \bibinfo {pages} {91} (\bibinfo {year} {1998})}\BibitemShut
  {NoStop}%
\bibitem [{\citenamefont {Prestipino}\ \emph {et~al.}(2012)\citenamefont
  {Prestipino}, \citenamefont {Laio},\ and\ \citenamefont
  {Tosatti}}]{prestipino:2012}%
  \BibitemOpen
  \bibfield  {author} {\bibinfo {author} {\bibfnamefont {S.}~\bibnamefont
  {Prestipino}}, \bibinfo {author} {\bibfnamefont {A.}~\bibnamefont {Laio}}, \
  and\ \bibinfo {author} {\bibfnamefont {E.}~\bibnamefont {Tosatti}},\ }\href
  {\doibase 10.1103/PhysRevLett.108.225701} {\bibfield  {journal} {\bibinfo
  {journal} {Physical Review Letters}\ }\textbf {\bibinfo {volume} {108}},\
  \bibinfo {pages} {225701} (\bibinfo {year} {2012})}\BibitemShut {NoStop}%
\bibitem [{\citenamefont {Lutsko}(2012)}]{lutsko:2012}%
  \BibitemOpen
  \bibfield  {author} {\bibinfo {author} {\bibfnamefont {J.~F.}\ \bibnamefont
  {Lutsko}},\ }\href@noop {} {\bibfield  {journal} {\bibinfo  {journal} {The
  Journal of chemical physics}\ }\textbf {\bibinfo {volume} {136}},\ \bibinfo
  {pages} {034509} (\bibinfo {year} {2012})}\BibitemShut {NoStop}%
\bibitem [{\citenamefont {Lutsko}\ and\ \citenamefont
  {Dur{\'a}n-Olivencia}(2013)}]{lutsko:2013}%
  \BibitemOpen
  \bibfield  {author} {\bibinfo {author} {\bibfnamefont {J.~F.}\ \bibnamefont
  {Lutsko}}\ and\ \bibinfo {author} {\bibfnamefont {M.~A.}\ \bibnamefont
  {Dur{\'a}n-Olivencia}},\ }\href@noop {} {\bibfield  {journal} {\bibinfo
  {journal} {The Journal of chemical physics}\ }\textbf {\bibinfo {volume}
  {138}},\ \bibinfo {pages} {244908} (\bibinfo {year} {2013})}\BibitemShut
  {NoStop}%
\bibitem [{\citenamefont {Schrader}\ \emph {et~al.}(2009)\citenamefont
  {Schrader}, \citenamefont {Virnau},\ and\ \citenamefont
  {Binder}}]{schrader:2009}%
  \BibitemOpen
  \bibfield  {author} {\bibinfo {author} {\bibfnamefont {M.}~\bibnamefont
  {Schrader}}, \bibinfo {author} {\bibfnamefont {P.}~\bibnamefont {Virnau}}, \
  and\ \bibinfo {author} {\bibfnamefont {K.}~\bibnamefont {Binder}},\ }\href
  {\doibase 10.1103/PhysRevE.79.061104} {\bibfield  {journal} {\bibinfo
  {journal} {Physical Review E}\ }\textbf {\bibinfo {volume} {79}},\ \bibinfo
  {pages} {061104} (\bibinfo {year} {2009})}\BibitemShut {NoStop}%
\bibitem [{\citenamefont {Jungblut}\ and\ \citenamefont
  {Dellago}(2015)}]{jungblut:2015}%
  \BibitemOpen
  \bibfield  {author} {\bibinfo {author} {\bibfnamefont {S.}~\bibnamefont
  {Jungblut}}\ and\ \bibinfo {author} {\bibfnamefont {C.}~\bibnamefont
  {Dellago}},\ }\href@noop {} {\bibfield  {journal} {\bibinfo  {journal} {The
  Journal of Chemical Physics}\ }\textbf {\bibinfo {volume} {142}},\ \bibinfo
  {pages} {064103} (\bibinfo {year} {2015})}\BibitemShut {NoStop}%
\bibitem [{\citenamefont {Shizgal}\ and\ \citenamefont
  {Barrett}()}]{shizgal:1989}%
  \BibitemOpen
  \bibfield  {author} {\bibinfo {author} {\bibfnamefont {B.}~\bibnamefont
  {Shizgal}}\ and\ \bibinfo {author} {\bibfnamefont {J.~C.}\ \bibnamefont
  {Barrett}},\ }\href@noop {} {\ \textbf {\bibinfo {volume} {91}},\ \bibinfo
  {pages} {6505}}\BibitemShut {NoStop}%
\bibitem [{\citenamefont {Sorokin}\ \emph {et~al.}(2017)\citenamefont
  {Sorokin}, \citenamefont {Dubinko},\ and\ \citenamefont
  {Borodin}}]{sorokin:2017}%
  \BibitemOpen
  \bibfield  {author} {\bibinfo {author} {\bibfnamefont {M.}~\bibnamefont
  {Sorokin}}, \bibinfo {author} {\bibfnamefont {V.}~\bibnamefont {Dubinko}}, \
  and\ \bibinfo {author} {\bibfnamefont {V.}~\bibnamefont {Borodin}},\
  }\href@noop {} {\bibfield  {journal} {\bibinfo  {journal} {Physical Review
  E}\ }\textbf {\bibinfo {volume} {95}},\ \bibinfo {pages} {012801} (\bibinfo
  {year} {2017})}\BibitemShut {NoStop}%
\bibitem [{\citenamefont {Kuipers}\ and\ \citenamefont
  {Barkema}(2010)}]{kuipers:2010}%
  \BibitemOpen
  \bibfield  {author} {\bibinfo {author} {\bibfnamefont {J.}~\bibnamefont
  {Kuipers}}\ and\ \bibinfo {author} {\bibfnamefont {G.}~\bibnamefont
  {Barkema}},\ }\href@noop {} {\bibfield  {journal} {\bibinfo  {journal}
  {Physical Review E}\ }\textbf {\bibinfo {volume} {82}},\ \bibinfo {pages}
  {011128} (\bibinfo {year} {2010})}\BibitemShut {NoStop}%
\bibitem [{\citenamefont {Kappler}\ \emph {et~al.}(2018)\citenamefont
  {Kappler}, \citenamefont {Daldrop}, \citenamefont {Brünig}, \citenamefont
  {Boehle},\ and\ \citenamefont {Netz}}]{kappler:2018}%
  \BibitemOpen
  \bibfield  {author} {\bibinfo {author} {\bibfnamefont {J.}~\bibnamefont
  {Kappler}}, \bibinfo {author} {\bibfnamefont {J.~O.}\ \bibnamefont
  {Daldrop}}, \bibinfo {author} {\bibfnamefont {F.~N.}\ \bibnamefont
  {Brünig}}, \bibinfo {author} {\bibfnamefont {M.~D.}\ \bibnamefont {Boehle}},
  \ and\ \bibinfo {author} {\bibfnamefont {R.~R.}\ \bibnamefont {Netz}},\
  }\href {\doibase 10.1063/1.4998239} {\bibfield  {journal} {\bibinfo
  {journal} {The Journal of Chemical Physics}\ }\textbf {\bibinfo {volume}
  {148}},\ \bibinfo {pages} {014903} (\bibinfo {year} {2018})}\BibitemShut
  {NoStop}%
\bibitem [{\citenamefont {Ruiz-Montero}\ \emph {et~al.}(1997)\citenamefont
  {Ruiz-Montero}, \citenamefont {Frenkel},\ and\ \citenamefont
  {Brey}}]{ruiz-montero:1997}%
  \BibitemOpen
  \bibfield  {author} {\bibinfo {author} {\bibfnamefont {M.~J.}\ \bibnamefont
  {Ruiz-Montero}}, \bibinfo {author} {\bibfnamefont {D.}~\bibnamefont
  {Frenkel}}, \ and\ \bibinfo {author} {\bibfnamefont {J.~J.}\ \bibnamefont
  {Brey}},\ }\href {\doibase 10.1080/002689797171922} {\bibfield  {journal}
  {\bibinfo  {journal} {Molecular Physics}\ }\textbf {\bibinfo {volume} {90}},\
  \bibinfo {pages} {925} (\bibinfo {year} {1997})}\BibitemShut {NoStop}%
\bibitem [{\citenamefont {Rein~ten Wolde}\ \emph {et~al.}(1996)\citenamefont
  {Rein~ten Wolde}, \citenamefont {Ruiz-Montero},\ and\ \citenamefont
  {Frenkel}}]{tenwolde:1996}%
  \BibitemOpen
  \bibfield  {author} {\bibinfo {author} {\bibfnamefont {P.}~\bibnamefont
  {Rein~ten Wolde}}, \bibinfo {author} {\bibfnamefont {M.~J.}\ \bibnamefont
  {Ruiz-Montero}}, \ and\ \bibinfo {author} {\bibfnamefont {D.}~\bibnamefont
  {Frenkel}},\ }\href@noop {} {\bibfield  {journal} {\bibinfo  {journal} {The
  Journal of chemical physics}\ }\textbf {\bibinfo {volume} {104}},\ \bibinfo
  {pages} {9932} (\bibinfo {year} {1996})}\BibitemShut {NoStop}%
\bibitem [{\citenamefont {Steinhardt}\ \emph {et~al.}(1983)\citenamefont
  {Steinhardt}, \citenamefont {Nelson},\ and\ \citenamefont
  {Ronchetti}}]{steinhardt:1983}%
  \BibitemOpen
  \bibfield  {author} {\bibinfo {author} {\bibfnamefont {P.~J.}\ \bibnamefont
  {Steinhardt}}, \bibinfo {author} {\bibfnamefont {D.~R.}\ \bibnamefont
  {Nelson}}, \ and\ \bibinfo {author} {\bibfnamefont {M.}~\bibnamefont
  {Ronchetti}},\ }\href@noop {} {\bibfield  {journal} {\bibinfo  {journal}
  {Physical Review B}\ }\textbf {\bibinfo {volume} {28}},\ \bibinfo {pages}
  {784} (\bibinfo {year} {1983})}\BibitemShut {NoStop}%
\bibitem [{\citenamefont {Kaestner}(2011)}]{kaestner:2011}%
  \BibitemOpen
  \bibfield  {author} {\bibinfo {author} {\bibfnamefont {J.}\ \bibnamefont
  {K\"astner}},\ }\href@noop {} {\bibfield  {journal} {\bibinfo  {journal}
  {Wiley Interdisciplinary Reviews: Computational Molecular Science}\ }\textbf {\bibinfo {volume} {1}},\
\bibinfo {pages} {932} (\bibinfo {year} {2011})}\BibitemShut {NoStop}%
\bibitem [{\citenamefont {Kaestner}(2005)}]{kaestner:2005}%
  \BibitemOpen
  \bibfield  {author} {\bibinfo {author} {\bibfnamefont {J.}\ \bibnamefont
  {K\"astner}},\bibinfo {author} {\bibfnamefont {W.}\ \bibnamefont
  {Thiel}},\ }\href@noop {} {\bibfield  {journal} {\bibinfo  {journal}
  {The Journal of Chemical Physics}\ }\textbf {\bibinfo {volume} {123}},\
  \bibinfo {pages} {144104} (\bibinfo {year} {2005})}\BibitemShut {NoStop}%
\bibitem [{\citenamefont {Mann}(1945)}]{mann:1945}%
  \BibitemOpen
  \bibfield  {author} {\bibinfo {author} {\bibfnamefont {H.B.}\ \bibnamefont
  {Mann}}\ }\href@noop {} {\bibfield  {journal} {\bibinfo  {journal}
  {Econometrica}\ }\textbf {\bibinfo {volume} {13}},\
  \bibinfo {pages} {245} (\bibinfo {year} {1945})}\BibitemShut {NoStop}%
\bibitem [{\citenamefont {Ingber}(1993)}]{ingber:1993}%
  \BibitemOpen
  \bibfield  {author} {\bibinfo {author} {\bibfnamefont {L.}\ \bibnamefont
  {Ingber}}\ }\href@noop {} {\bibfield  {journal} {\bibinfo  {journal}
  {Mathematical and Computer Modelling}\ }\textbf {\bibinfo {volume} {18}},\
  \bibinfo {pages} {29} (\bibinfo {year} {1993})}\BibitemShut {NoStop}%
\end{thebibliography}
\end{document}